\documentclass{aa}
\usepackage{graphicx}
\usepackage{txfonts}
\usepackage{epstopdf}
\usepackage{supertabular}
\begin{document}

\title{Recent star formation history of the Large and Small Magellanic Clouds}
\author{Indu, G.\inst{1}, Annapurni Subramaniam\inst{1}}
\institute{Indian Institute of Astrophysics, Koramangala II Block, Bangalore-560034, India\\
           \email{indu@iiap.res.in, purni@iiap.res.in}}
\date{Received, accepted}
\abstract
{}
{ }
{We traced the age of the last star formation event (LSFE) in the inner Large \& Small Magellanic Cloud (L\&SMC) using the photometric data in V and I pass bands from the Optical Gravitational Lensing Experiment (OGLE-III) and the Magellanic Cloud Photometric Survey (MCPS). The LSFE is estimated from the
main-sequence turn off point in the color-magnitude diagram (CMD) of a sub-region. After correcting for extinction, the turn off magnitude is converted to age, which represents the LSFE in a region.}
{The spatial distribution of the age of the LSFE shows that the star formation has shrunk to the central regions in the last 100 Myr in both the galaxies. The location as well as age of LSFE is found to correlate well with those of the star cluster in both the Clouds.
The SMC map shows two separate concentrations of young star formation, one near the center and the other near the wing. We detect peaks of star formation at 0 - 10 Myr, 90 - 100 Myr in the LMC, and 0 - 10 Myr, 50 - 60 Myr in the SMC.
The quenching of star formation in the LMC is found to be asymmetric with respect to the optical center such that most of the young star forming regions are located to the north and east. On deprojecting the data on the LMC plane, the recent star formation appears to be stretched in the north-east direction and the HI gas is found to be distributed preferentially in the North. The centroid is found to shift to north in 200 - 40 Myr, whereas it is found to shift to north-east in the last 40 Myr. In the SMC, we detect a shift in centroid of the population younger than 500 Myr and up to 40 Myr in the direction of the LMC.}
{ We propose that the HI gas in the LMC is pulled to the north of the LMC in the last 200 Myr due to the gravitational attraction of our Galaxy at the time of perigalactic passage. The shifted HI gas is preferentially compressed in the north during 200 - 40 Myr and in the north-east in the last 40 Myr, due to the motion of the LMC in the Galactic halo. The recent star formation in the SMC is due to the combined gravitational effect of the LMC and the perigalactic passage.}

\authorrunning{Indu \& Subramaniam}
\titlerunning{Star formation history of the Magellanic Clouds}
\maketitle

\section{Introduction}
The Large and Small Magellanic Clouds (LMC \& SMC) along with the components, the bridge and the stream, comprises the Magellanic system. The presence of the bridge and the stream connecting the two Clouds suggest that these two galaxies might have been together possibly as an interacting pair. The Bridge in
particular indicates that they have had a close encounter in the
recent past. This system moves in the gravitational potential of the Galaxy. It is obvious that the structure, kinematics and evolution of the clouds and the Galaxy are modified by their interactions. The Magellanic Clouds are gas rich and have active on-going star formation, possibly triggered due to interactions between the Magellanic Clouds (MCs) and with the Galaxy. It was long believed that the Clouds orbit our Galaxy and the bursts of star formation episodes seen in both the Clouds are probably due to their perigalactic passage and the tidal effects (\cite{hz04}, \cite{ljk95}). On the other hand, the recent estimates of the proper motion of the Clouds find that the Magellanic System is probably passing close to the Milky Way (MW) for the first time (\cite{bk07}). Thus the star formation episodes which were attributed due to the perigalactic passage will need to be reconsidered. Nevertheless, the star formation
history (SFH) of the MCs were studied to identify the interaction between the Clouds and their ages, assuming that an interaction will induce a simultaneous star formation in both the
galaxies. Also, the star formation induced due to interaction can produce propagating star formation in the galaxies, the direction of propagation could give valuable clues to the details of interaction. Since the Clouds are passing near the Galaxy presently and are together, the pattern of the recent star formation is likely to indicate the effects due to the Galaxy-LMC-SMC interaction. In this work, we plan to study the pattern of the recent star formation in the Clouds, with specific interest to trace the origin and nature of the interaction which caused it.

The recent SFH has been studied by various authors using star clusters as well as 
field star population. The star clusters in the LMC are studied and their derived age distribution is compared with that in the SMC by \cite{pu00}. The comparison of both cluster formation and star formation is also done to find the correlation between the two processes(\cite{h99}, \cite{s04}). \cite{hz09} (hear after H$\&$Z09) presented a reconstruction of SFH of the LMC  with a conclusion that field and cluster star formation modes are tightly coupled. They found a quiescent epoch from 12 to 5 Gyr ago and star formation peaks at 2 Gyr, 500 Myr, 100 Myr $\&$ 12 Myr. The spatial distribution of cluster as well as star formation rates also form a regime of interest. An investigation on the bar cluster population in the LMC (\cite{b92}) has shown that clusters younger than 200 Myrs are not homogeneously distributed through the bar. In particular, a strong star formation event at 100 Myr was detected in the eastern part of the bar. \cite{hz04} studied the SFH of the SMC and found a 
quiescent epoch between 8.4 to 3 Gyr. They also found a continuous star formation from 3 Gyr to the present epoch with star formation peaks at 2-3 Gyr, 400 \& 60 Myr. \cite{n09} also studied the SFH of the SMC and found that the younger stars (200 - 500 Myrs) have an asymmetric distribution with the appearance of the wing, while the older population ($>$ 1 Gyrs) is distributed similarly at all radii and all azimuth. In contrast with \cite{hz04}, \cite{n09} did not find any quiescent epoch in the intermediate ages. 
\cite{g10} studied the SFH of both the clouds based on star clusters with age $<$1 Gyr. They found that the cluster formation peaks at 160 and 630 Myr for the SMC $\&$ 125 and 800 Myr for the LMC. Thus, the studies done so far has found that the age of the star formation peaks in the LMC and the SMC fall in the similar range, but the values do not coincide. \cite{a99} found that there is a propagating star formation in the last 100 Myr, along the bar, from southeast to northwest, detected using MACHO Cepheids.
Though some studies have found that there is evidence for propagating star formation, the details are not clear. Study of a larger area using homogeneous data is required to obtain the details of any propagating star formation. 

There are different models for the evolution of star formation in irregular - low mass spiral galaxies which depicts spatial variation of star formation. \cite{g08} presented an out side in disk evolution in the LMC which is explained in terms of decreasing HI column density with galactocentric distance. They observed an outside to inside quenching of star formation, such that, a field at 2.3$^{\circ}$ is currently active in star formation, while fields at 4.4$^{\circ}$, 5.5$^{\circ}$, and 7.1$^{\circ}$, have 100 Myr, 0.8 Gyr and 1.5 Gyr old stars as the youngest population respectively. Thus, the age of the youngest stars in each field gradually increases with the distance from the center and the population is found to be older on an average towards the outer part of the galaxy (\cite{s10}, \cite{g09}). It will be interesting to see up to what inner radius this outside in quenching of star formation can be traced. Such a study also requires homogeneous data over a large spatial area of the two Clouds. 

The most common method used for the estimation of SFH involves a quantitative comparison of observed colour-magnitude diagrams (CMD) with synthetic CMDs constructed using theoretical isochrones according to an adopted IMF and input model SFH. H$\&$Z09 have performed this for the LMC using the MCPS data, where they have modeled the complete SFH of the LMC. They used just five time steps for the recent SFH (6.3 Myr, 12.5 Myr, 25 Myr, 50 Myr and 100 Myr), which is equal to 0.3 in log(age). The gaps between the steps are large and these large gaps in the recent SFH does not give enough time resolution to identify and study any
propagating star formation. If one aims to study the recent SFH with a time resolution of 5-10 Myr for ages less than 100 Myr, 
then it is better to model only the young stars and not the entire range of stars in a given region. Since large numbers of stars are required in the CMD of a region to model the full range of age, the area required is also large which reduces the spatial resolution. In order to achieve a better spatial as well as temporal resolution, we have adopted a different method. In this method, we estimate the age of the last star formation event (LSFE) in a given region, by identifying the turn-off of the main sequence (MS) in the CMD of the corresponding region. The turn-off identified from the luminosity function (LF) of the MS represents the last star formation experienced by the region. The reddening to the region is also estimated from the turn-off. The spatial map of age of the LSFE is used to identify the presence of any propagating star formation. We also produce a map of the average reddening in regions studied in both the galaxies. The spatial and temporal resolution achieved using this method are better than those obtained using the traditional method, as only a small range in age is studied. For the same reason, this method does not assume any age metallicity relation. 

The paper is organised as follows: Section 2 deals with Data and section 3 deals with Methodology. The results are presented in section 4, with subsections for extinction and LSFE maps of the LMC and the SMC. Discussion is presented in section 5. The error estimates are presented in appendix.

\section{Data}
This study makes use of two publically available photometric survey data which cover large area of the MCs. These are the catalogs produced by the Optical Gravitational Lensing Experiment (OGLE-III) and the Magellanic Cloud Photometric Survey (MCPS).
OGLE III photometric maps form a significant extension of OGLE II. The total observed area is 40 square degrees, covering 116 LMC regions, each of which covers an area of 35 x 35 arcmin$^2$. The survey presents mean calibrated photometry in V and I pass bands of about 35 million stars. Each of the 116 regions are observed using 8 chips each having an area of 8.87 x 17.74 arcmin$^2$. In this study, each of these are divided into sub regions of different areas 4.43 x 4.43, 4.43 x 8.87 $\&$ 8.87 x 8.87 arcmin$^2$. The different sizes for the regions are chosen to study the effect of area on the identified turn- off magnitude and thus in estimating the age of the LSFE. For the SMC the survey presents VI photometry of 6.5 million stars from 41 fields, covering an area of 14 square deg in the sky. The area binning in SMC is done with  4.43 x 4.43, 4.43 x 8.87 $\&$ 8.87 x 8.87 arcmin$^2$, for this study.

MCPS spans a total region of 64 square deg, with about 24 million objects in the LMC. The survey presents photometric as well as extinction maps in U, B, V $\&$ I pass bands. In the case of the MCPS, the subregions have area in the range,  10.5 x 30 arcmin$^2$, 10.5 x 15 arcmin$^2$ $\&$ 5.3 x 15 arcmin$^2$. The MCPS survey presents UBVI photometry of the central 18 square deg of the SMC covering 5 million stars. Subregions of area 10.5 x 30 arcmin$^2$, 10.5 x 15 arcmin$^2$ $\&$ 5.3 x 15 arcmin$^2$ were made to estimate the age of the LSFE. For uniformity, we use the V and I photometric data from both the catalogs.
 
\section{Methodology}
\subsection{Identifying the MS Turn-off}
We have adopted the following method to identify the age of the LSFE from the CMDs. The observed region is divided into several smaller sub-regions to increase spatial resolution. The area for the smallest sub-region is decided based on the number of MS stars in the CMD, that is, it has a minimum of about 250 MS stars in the SMC \& 600 MS stars in the LMC.
For each sub-region, (V$-$I) vs V CMD is constructed and the MS is identified as stars brighter than 21 mag and with a color index less than 0.5 mag. The turn off is identified from the MS by constructing a luminosity function (LF) by binning in V magnitude with a bin size of 0.2 mag. The brightest bin in the LF is identified using a statistical cut off of 2$\sigma$ significance (minimum 5 stars in the brightest bin). The cut off
means that the tip of the MS should have atleast 5 stars such that it has a variance of $\sqrt(5)=2.2$. Thus, the number of stars identified as the tip of the MS is more than twice the variance. This cut-off is chosen in order to reduce the statistical fluctuation in identifying the MS turn-off. Thus, after computing the LF of the MS, the brightest bin which has atleast 5 or more stars is found and this is identified as the tip of the MS.  This condition also implies that, the age of the LSFE identified in a sub-region will have a certain threshold star formation rate to form these many stars. On the other hand, star formation events with rates lower than this threshold will not be identified. The above condition would also minimise the chances of identifying blue super giants as MS stars. Since the number of stars in the brightest bin is dependent on the area used, the age of the LSFE will also depend on the area of the bin considered. Thus, we have used three sizes for the area of the sub-regions, to map the age of the LSFE.   The average V magnitude corresponding to the brightest bin with the required number of stars is taken as the tip of the MS. This is considered as the turn-off magnitude $V_{to}$ of the youngest stellar population present in the region.

In order to convert $V_{to}$ to age, it has to be corrected for extinction. This is estimated from the colour of the turn-off. In general, colour of the turn-off is identified as the densest point on the MS, which will appear as a peak in a colour distribution of the stars near the turn-off. In other words, the peak of the distribution of stars with respect to (V$-$I) color near the turn-off can be 
used to identify the colour of the MS. In order to estimate the peak (V$-$I) color of the turn-off, a strip parallel to (V$-$I) axis with a width 0.5 magnitude is considered ($V_{to}$ + 0.5 mag). This strip is binned in colour (with a bin size of 0.1 mag) to study the distribution of stars along the (V$-$I) colour. This distribution is found to have a unique peak and asymmetric wings. The (V$-$I) bin corresponding to the peak of the distribution is identified as the location of the MS. The average value of the bin 
corresponding to this peak is taken as the (V$-$I) colour of the MS turn-off. Since we consider the peak
of the distribution and not the average, the colour estimated corresponds only to the MS stars. As the distribution is found to be asymmetric, the average is likely to be redder than the peak. 
The bin sizes used to estimate the turn-off magnitude and colour are same for both the LMC \& the SMC. Since we statistically trace the brightest part of the CMD in a region, this method is not much affected by crowding or incompleteness in the data. The following section describes the method to estimate the reddening and extinction.

\subsection{Estimation of Reddening and turn-off age}
We have estimated the reddening to a sub-region, from the colour of the turn-off identified from the CMD. The reddening is estimated as the difference between the estimated colour of the turn-off and the expected colour. Therefore one  requires to know the expected M$_v$ and the (V$-$I)$_0$.  Since we know the distance modulus (DM), we can approximately estimate the M$_v$ by applying an average value of extinction to the observed apparent magnitude. An initial extinction value is needed to calculate the M$_v$.
\begin{center}
$M_v = m_v- DM -A_v$
\end{center}
where $m_v$ is the apparent magnitude.
To begin with, in the case of the LMC, we assumed  an extinction of A$_v$$=$ 0.55 (\cite{z04}), along with a distance modulus of 18.5. This value of A$_v$ is the average value of the extinction towards the LMC. M$_v$ thus obtained is used to identify the approximate location of the turn-off and its expected color $(V-I)_0$ from the isochrones 
of (\cite{m08}) for a metallicity of Z=0.008. The difference
between the expected colour and the observed colour of the turn-off is defined as the colour excess for a sub-region.
\begin {center}
						$E(V$-$I) = (V$-$I) - (V$-$I)_0$
\end {center}

Then the actual $A_v$ for each subregion is found using the formula, (\cite{n04})
\begin {center}
                                                           
						$A_v = 2.48 E(V$-$I)$ 
\end {center}
  
Thus, the reddening, E(V$-$I) and extinction, A$_v$ are estimated for the LMC subregions. In the case of the SMC, a constant distance modulus of 18.9 and an average extinction value of $A_v$ $=$ 0.46 mag (\cite{z02}) are used. The expected value of the unreddened colour, $(V-I)_0$ is obtained from \cite{m08} isochrones for a metallicity of Z=0.004.  The above equations are used to estimate the extinction towards each subregion. Using these estimated values of extinction, actual $M_v$ corresponding to the turn-off is estimated for all the subregions.
Applying the calculated colour excess and extinction for each subregion, CMDs can be obtained with absolute magnitude, M$_v$ and dereddened color, (V$-$I)$_0$.  CMDs of some sample regions are shown in figure 1. The upper panels show the CMDs for the LMC subregions (MCPS on the left and OGLE III on the right), whereas the lower panels show the CMDs for the SMC subregions. The red dot on the MS shows the location of the identified turn-off. We have shown various cases here, where we can see broad/tight MS turn-off. 
 For example, consider the region shown in the top right panel, which corresponds to the location with mean RA = 79.8$^o$ \& Dec = $-$69.3$^o$. It covers an area of 4.43 x 8.87 arcmin$^2$. The total number of stars are about 6090. The identified apparent turn off  magnitude is 16.5, and the color index is $-$0.05. Applying the initial extinction value and DM as described above, the approximate location of the MS is identified to have an absolute magnitude of $16.5 - 18.5 - 0.55 = -2.55$. The nearest absolute $M_v$ value from the isochrone table is $-2.54$. The corresponding $(V - I)_0$ value is $-0.287$. Thus, the reddening towards this region is estimated as $E(V-I) = 0.237$ \& a corresponding extinction, $A_v = 0.588$ mag. Correcting the turn off $m_v$ for extinction using these values, we estimate the MS turn-off as $M_v$ as $-2.588$ mag. Converting this turn off magnitude to age with the estimated conversion relation, we get $log(age) = 7.41$.

The estimated turn-off $M_v$ is used to estimate the age of the LSFE in each sub-region. We estimated an age-M$_v$ relation for the turn-off, using the isochrones of \cite{m08}. The age-M$_v$ relation is obtained for both the LMC and the SMC separately, where the metallicity of the isochrones is chosen to be 0.008 for the LMC and 0.004 for the SMC. The plot of the log (age) vs M$_v$ for the L\&SMC are given in figure 2. The relation is found to be linear and we have derived
a linear relation between the two by fitting a line to the data points. The turn-off ages for the sample regions, estimated using the above relation are also shown in figure 1. The identification of MS turn-off
was found to be ambiguous for turn-off fainter than 18.0 magnitude. Therefore, we have put a limiting apparent turn-off magnitude of 18.0, for both the L\&SMC, which will eventually lead to the higher cut-off age of the LSFE, around 120 Myr for the LMC and 100 Myr for the SMC. Therefore, sub-regions with turn-off magnitude fainter than 18.0 magnitude are not considered and these locations will appear as gaps in the extinction and LSFE maps.

 Since the number of stars in the CMD increases with the area of the subregion considered, the derived turn-off parameters, extinction and the age of the LSFE will depend on the area considered. In order to understand the effect of area on the estimated parameters, we have derived the extinction and the age of the LSFE for all the three sizes of subregions, in both the data sets and in both the galaxies.  We have estimated the
errors in the estimation of extinction and age using two methods. We have simulated synthetic CMDs using
\cite{m08} isochrones for the LMC and the SMC. The synthetic CMDs are analysed similarly to the observed CMDs to estimate the errors as a function of the turn off magnitude and colour, extinction and sampling. The errors are also estimated using the method of propagation of errors. These are presented in the appendix.

%% Figure 1 (CMDs)

\begin{figure}
%\\begin{minipage}{152mm}
   \resizebox{\hsize}{!}{\includegraphics{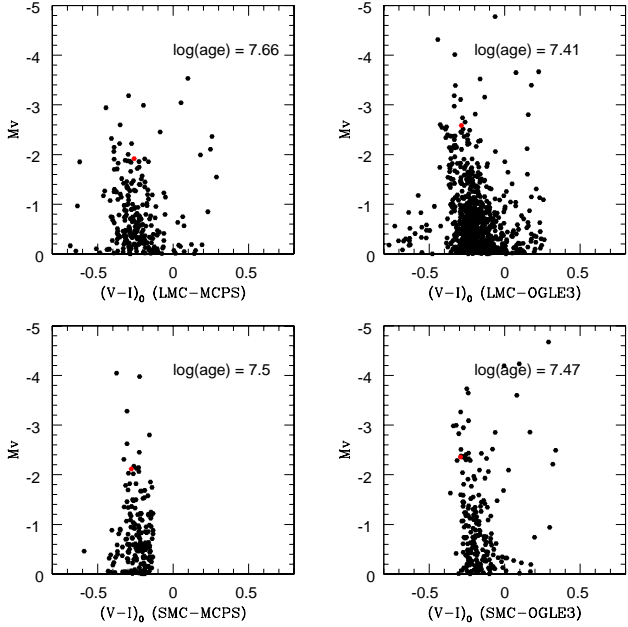}}
   \caption{ $M_v$ vs $(V-I)_0$ CMDs for two regions each in the L\&SMC. The top panels show two regions in the LMC, with left panel using MCPS data \& right using OGLE III data. Similarly bottom panels show two regions in the SMC (left using MCPS and right using OGLE III). The red dot 
marks the turn-off point, and the estimated turn off age is also shown.}
    %\\end{minipage}
    \end{figure}
    
%% Figure 2  (Age vs Mv relation)
\begin{figure}
%\\begin{minipage}{152mm}
   \resizebox{\hsize}{!}{\includegraphics{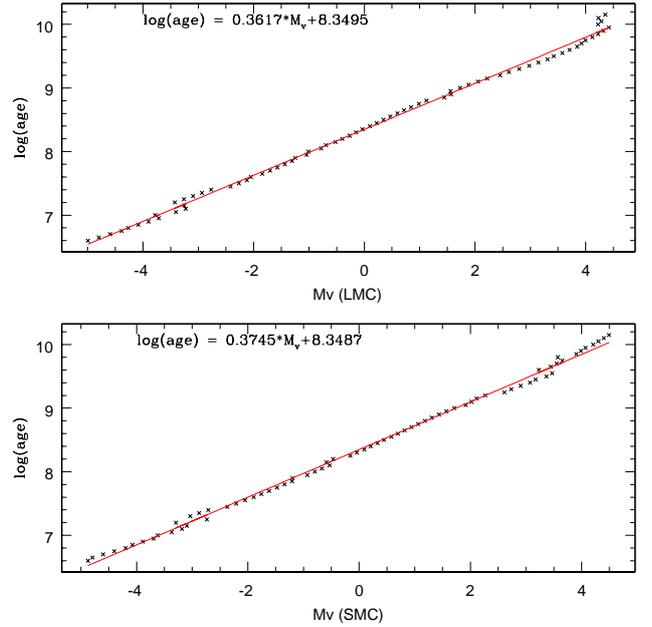}}
   \caption{The log(age) vs M$_v$ plots for L\&SMC, in the left \& right panels respectively. The turn-off M$_v$
corresponding to various ages are taken from \cite{m08} isochrones. The fitted line is shown in red in both the plots.}
    %\\end{minipage}
    \end{figure}
%% Figure 3 (Reddening map- MCPS)
\begin{figure}
%\\begin{minipage}{156mm}
   \resizebox{\hsize}{!}{\includegraphics{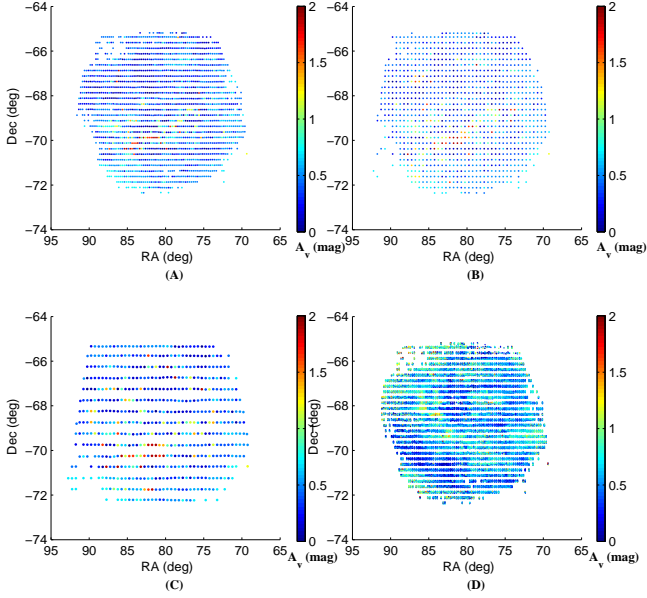}}
   \caption{The Extinction map of the LMC using MCPS data in the RA-Dec plane with three different area binning, A. 5.3 x 15 arcmin$^2$, B. 10.5 x 15 arcmin$^2$ $\&$ C. 10.5 x 30 arcmin$^2$. The bottom right panel is the extinction map of \cite{z04}, here the regions used in our analysis are selected and shown. Color coding is according to the $A_v$ value, which varies from 0.2 to 2.0 as shown in the color bar.}
    %\\end{minipage}
    \end{figure}
%% Figure 4 (Reddening map - OGLE III)
\begin{figure}
%\\begin{minipage}{156mm}
   \resizebox{\hsize}{!}{\includegraphics{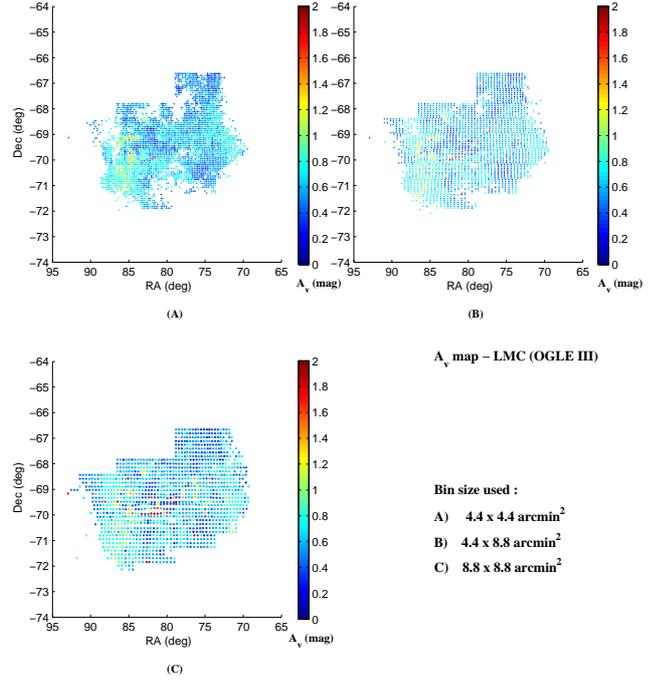}}
   \caption{The Extinction map of the LMC, similar to figure 3, using OGLE III data.}
    %\\end{minipage}
    \end{figure}
%% Figure 5 (Histogram) (LMC(MCPS) ours and Zaritsky)
\begin{figure}
%\\begin{minipage}{152mm}
   \resizebox{\hsize}{!}{\includegraphics{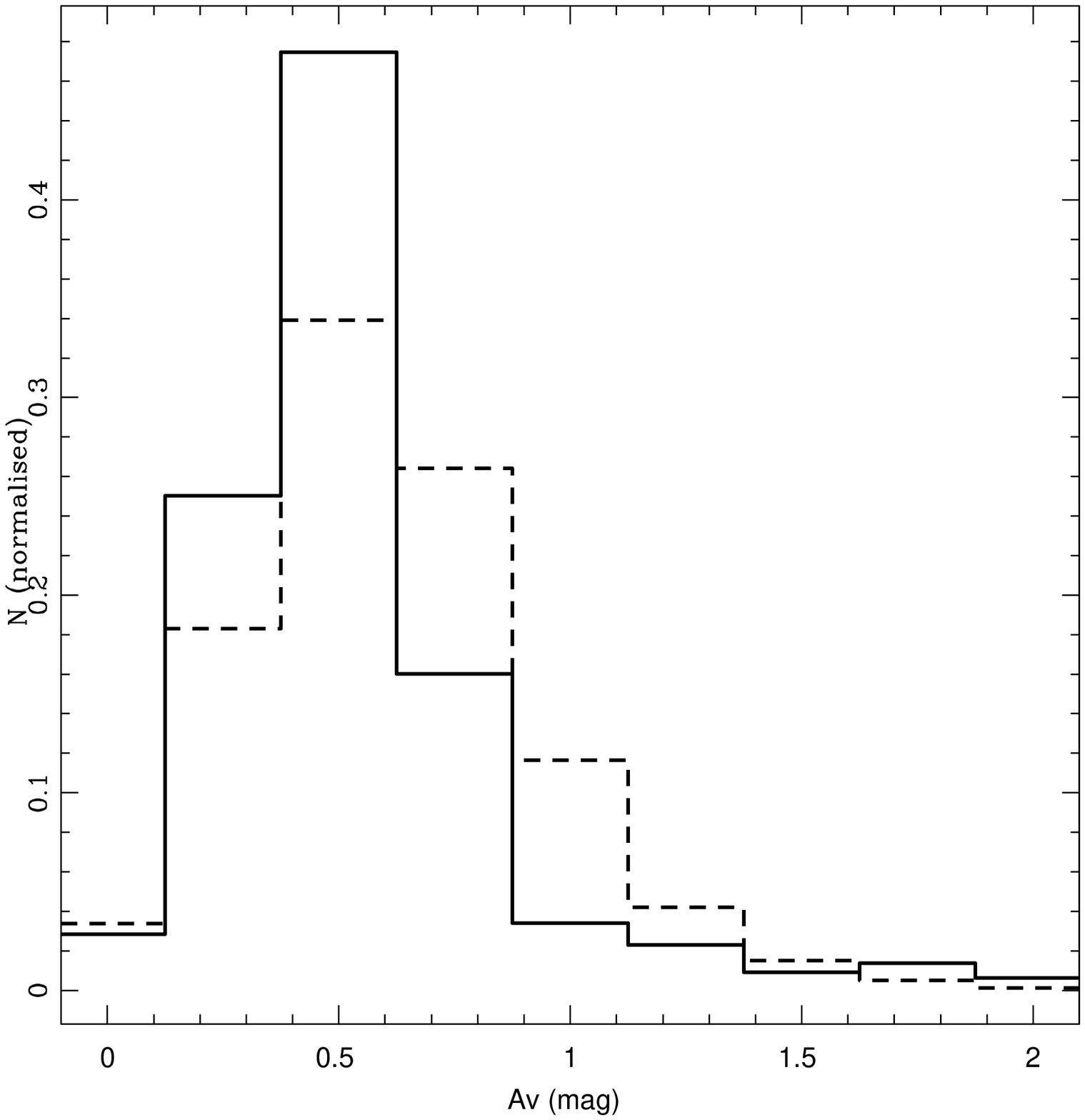}}
   \caption{The estimated distribution of extinction
	    in the LMC (shown as solid line) is compared with  distribution obtained from the extinction map of hot stars provided by \cite{z04} (shown as dashed line, extracted from the fits image from the website).}
    %\\end{minipage}
    \end{figure}
\section{Results}
\subsection{LMC: Extinction}
The estimated colour excess, E(V$-$I) and extinction, $A_v$ for all the regions can be used to create an extinction map of the LMC. The extinction is estimated from the brightest MS stars in each region. Figure 3 shows the map estimated using the MCPS data, and figure 4 shows the map estimated using the OGLE III data. Each figure has three panels and these show the extinction estimated using area bins as indicated in the figure. Since the estimated extinction would depend on the area used, the corresponding maps are used for the estimation of age of the LSFE.

The extinction map as estimated from the MCPS data (figure 3) shows that the extinction varies between $A_V$ = 0.2 --2.0 mag. Relatively large extinction is seen in the bar region, with the eastern 
part having higher extinction. A few regions in the north eastern part are also found to show large extinction. These features are seen in all the three plots, which show extinction for different area bins. It can be seen that with the increase in area, the estimated extinction increases. The
average value increases from about 0.4 to 0.6 mag, from the maps A to C. 
The extinction map presented in figure 4 is obtained from OGLE III data and has lesser area coverage and higher resolution. The extinction estimated here also ranges between $A_V$ = 0.2 - 2.0 mag. These maps show that the bar region has the highest extinction, along with the 30 Doradus region. The northern star forming regions are not covered here. The eastern region is found to have large extinction and this region 
coincides with the location of massive HI clouds, extending up to 30 Doradus star forming region. The map A
shows that the extinction has a very clumpy distribution with pockets of lesser extinction. We notice
the increase in the extinction with the increase in the area in these maps also. These maps will be useful for studies related to young stars, since the reddening in the LMC is known to
depend on the population studied (\cite{z04}). 

\cite{z04} have published a reddening map of the
early type stars in the LMC derived from the MCPS data using stars with effective temperature 12000 to 45000K. They have presented individual reddenings to 
stars and hence can be considered as a high resolution map. We compare the reddening distribution derived in this paper with that of \cite{z04}.  The histograms in figure 5 shows the 
distribution of our extinction values (shown as solid line) and the extinction map provided by \cite{z04} (shown as dashed line). We have used the 10.5 x 15 arcmin$^2$ area bin for comparison.
The distributions are found to be more or less similar. The peaks of both the distribution coincide at A$_v$ = 0.5 mag. The reddening estimated here in this study has more regions with less than the peak value. 
This may be because of the fact that we estimate extinction for stars located on the MS and ignore stars which are redder than the MS. Since we use a bin size of 0.2 mag in colour, the mean of this bin is taken as the colour of the MS and hence could introduce a shift of $\pm$0.1 mag in reddening. This will correspond to a shift of 0.25 in extinction. The variation one notices between the two distributions, as seen in figure 5, is of this order. This variation in the extinction measurement can introduce a shift in the estimated age and this will be included in estimating the error in age estimation (see appendix).  In figure 3 D, we have shown extinction map as estimated by \cite{z04} for regions which are analysed in this study. This map was obtained by spatially correlating the \cite{z04} map with our sample of selected regions in the MCPS data. The map includes extinction estimates of all the stars present in a given region. This map can be compared with
the MCPS extinction maps derived in this study (shown in figure 3a - 3c). The maps are more or less comparable. We do not detect isolated high extinction
values, probably because we derive extinction values for regions and not individual stars.
% but the  distribution shifted by 0.2 magnitude towards the higher values. Thus, the extinction estimated here are found to be lesser by 0.2 mag. This could be due to the fact the brightest and the reddest stars are not used to estimate the reddening, as only those identified as located near the turn-off are used. 
%Since the age is estimated using the turn-off, this value of reddeing
%is relevant for the age estimation. A shift of 0.1 in A$_v$ will give rise to an age shift of log(age) equal to 0.072 and 0.075 for the LMC and the SMC respectively.
%% Figure 6new (Reddening map - H&Z)
%\begin{figure}
%%\\begin{minipage}{156mm}
 %  \resizebox{\hsize}{!}{\includegraphics{extnhz.eps}}
 %  \caption{The Extinction map of the LMC from (\cite{z04}).}
 %   %\\end{minipage}
  %  \end{figure}
%Figure 6 (LSFE map -MCPS)
\begin{figure}
%\\begin{minipage}{156mm}
   \resizebox{\hsize}{!}{\includegraphics{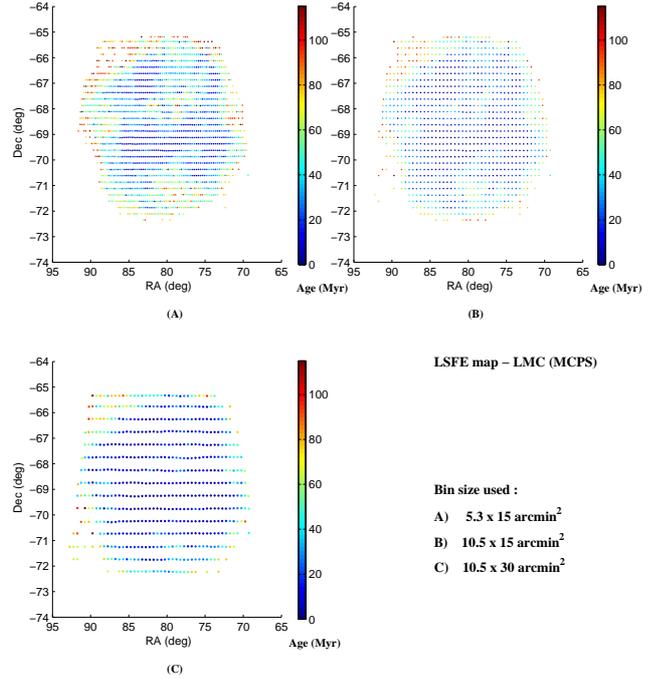}}
   \caption{The LSFE map of the LMC using MCPS data, in the RA-Dec plane with three different area binning, as specified in the figure. Color coding is according to the LSFE age as shown in the color bar. }
    %\\end{minipage}
    \end{figure}
%% Figure 7 (LSFE map - OGLE III)
\begin{figure}
%\\begin{minipage}{156mm}
   \resizebox{\hsize}{!}{\includegraphics{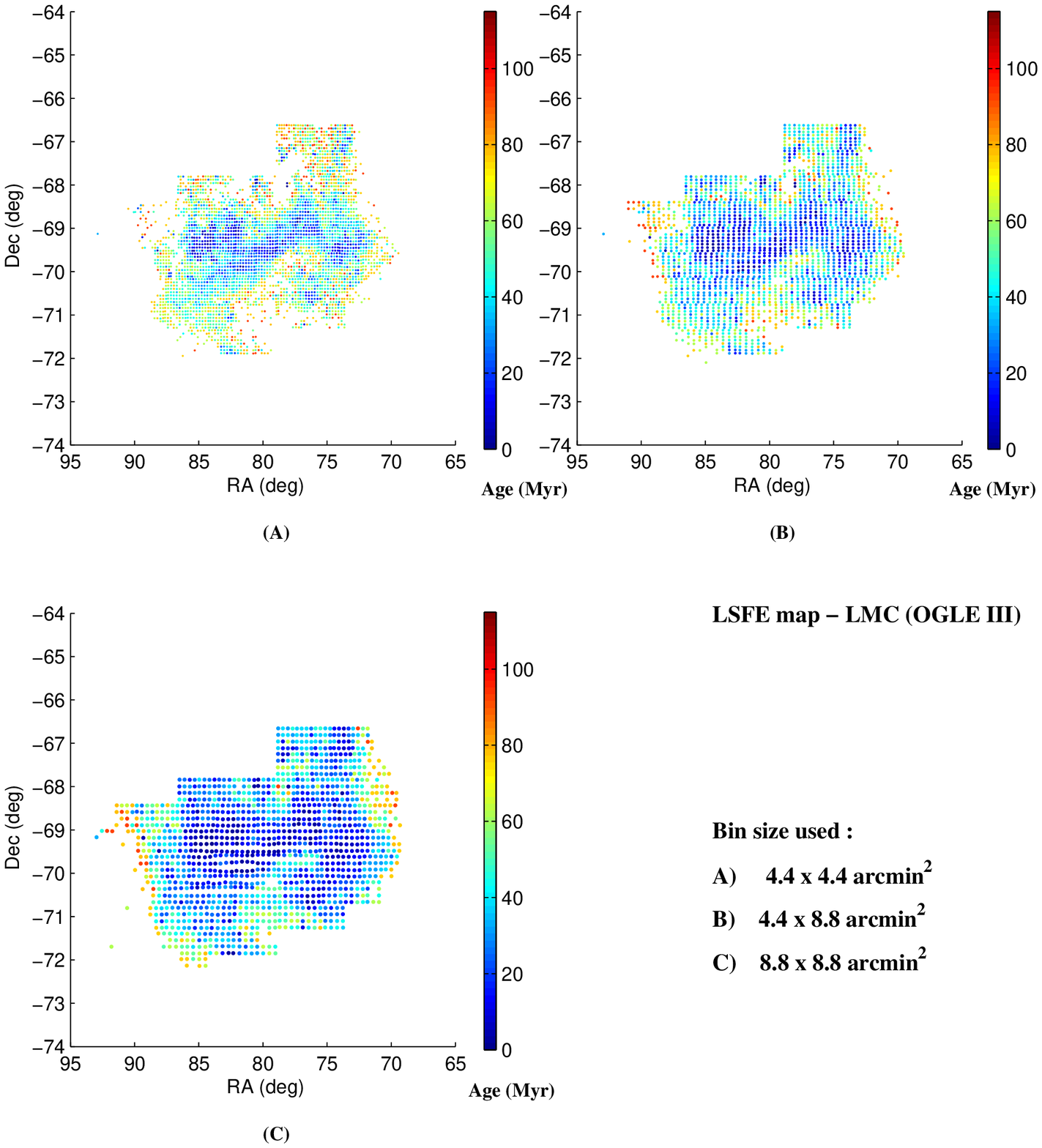}}
   \caption{The LSFE map of the LMC, similar to figure 6, using OGLE III data.}
    %\\end{minipage}
    \end{figure}\\
%% Figure 8 ( LMC age histogram)
 \begin{figure}
%\\begin{minipage}{152mm}
   \resizebox{\hsize}{!}{\includegraphics{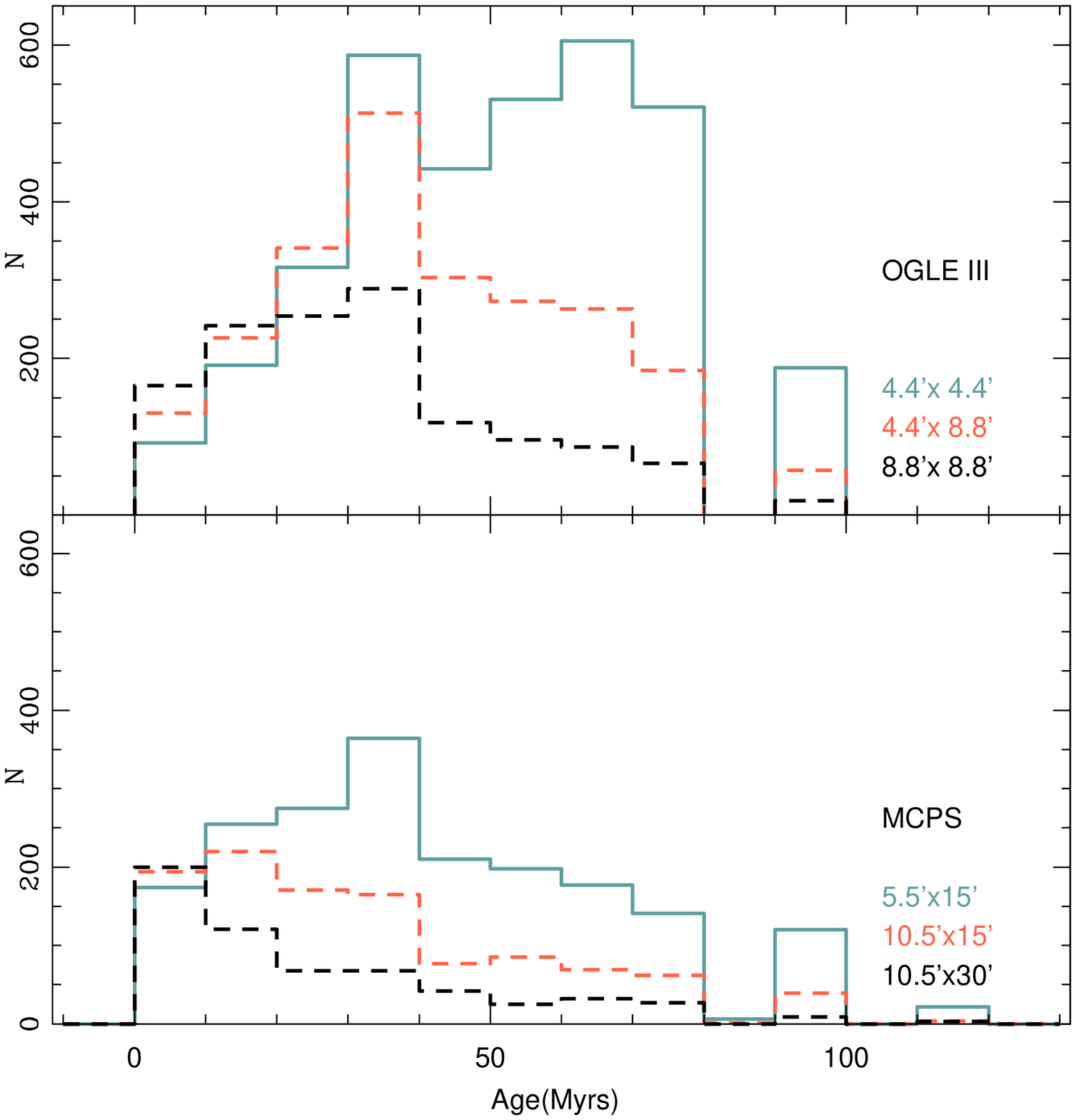}}
   \caption{The statistical distribution of LSFE ages of the LMC. Upper panel shows OGLE III data and 
lower panel shows MCPS data, with the three colors corresponding to different area, as specified in the figure.}
    %\\end{minipage}
    \end{figure}

%% Figure 9 ( LSFE MCPS map - xy plane, rings)
\begin{figure}
%\\begin{minipage}{152mm}
   \resizebox{\hsize}{!}{\includegraphics{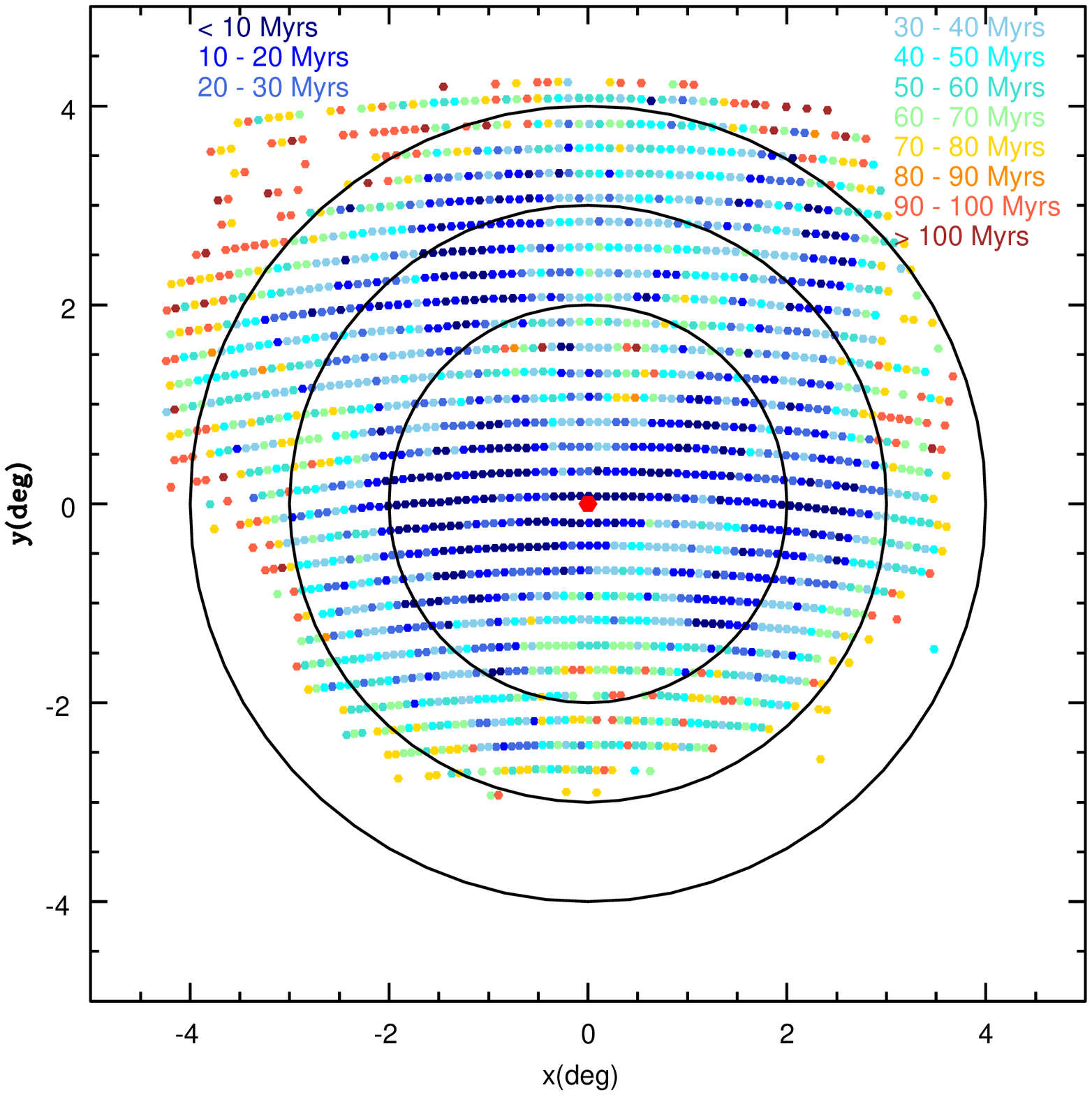}}
   \caption{The LSFE map of the LMC(same as A in figure 6) in the projected x - y plane of the sky. Three concentric rings of radii 2$^o$, 3$^o$, \& 4$^o$ are over plotted.The red dot at (0,0) is the optical center of the LMC.}
    %\\end{minipage}
    \end{figure}

\subsection{LMC: Distribution of age of the LSFE}
We have estimated the age of the LSFE across the LMC for subregions of three different sizes.
The distribution of age of the LSFE can be used to study 
the quenching as well as propagation of star formation in the central regions of the LMC. The maps of the age of the LSFE derived from the MCPS and the OGLE III data are shown in figures 6 and 7 respectively. 

Figure 6 shows that the age of the LSFE ranges from 0-120 Myr. The figure shows maps for three different area bins and these are indicated. When we inspect the
high resolution map A, we can see that it is a clumpy distribution with many central pockets having
very young age, suggested by the dark blue colour points. The youngest ages are found in the bar region, 
near the 30 Doradus and the northern regions. The locations are similar to the regions identified
by H\&Z09, such as the blue arm, constellation III and 30 Doradus. These young pockets are surrounded by regions with older star formation.
We can also see that the age of the LSFE is found to progressively increase when we go
towards the outer regions.  We also see a couple of small pockets of older
star formation in the inner regions. The map also brings out the north-western void. On the whole, the inner regions have ages in the range 0 -40 Myr, whereas, regions towards the periphery have ages in the range, 60 -100 Myr. The age map for larger area
bins are shown in B and C. These maps bring out lesser details, when compared to the map A. The clumpyness 
seen in map A disappears in B and C. As mentioned earlier, the age limit of the technique is about 120 Myr and we have not considered regions which have older turn-off. These regions are expected to appear as gaps in the map. It can be seen that these gaps do not appear in the inner regions, whereas it appears towards the periphery. In the case of map A, the outer regions are missing due to the above reason. If we inspect maps B and C, we can see that more and more outer regions are covered in these maps. This is due to the fact that outer regions have older turn-off and their ages become younger when area becomes larger. The point to be noted is that most of the regions in the inner 3$^o$ have experienced star formation in the last 100 Myr.
Another aspect brought out by these maps are that the age of the LSFE gets older with radius. The average
age is around 20 Myr in the central regions, whereas it is about 80 Myr near the periphery. These values
are found to be more or less similar for all the three area bins. 

Figure 7 shows the age map of the LSFE as estimated from the OGLE III data. Map A has the highest spatial
resolution and it shows a clumpy distribution of ages. The central region, which includes the bar 
region and the 30 Doradus region are found to have young ages. These sites have continued to form stars till very recently. Similar to the MCPS map, we see that the star forming regions have shrunken to smaller
and smaller pockets, where the star formation still continues. These pockets are the bar region and the 30 Doradus region. The spiral type pattern one could see in the western side of the bar in figure 6, is found to break into smaller multiple regions with star formation. Since the OGLE III has lesser coverage of the north, we are unable to study the northern star forming regions. We see a gradual increase in the age of the LSFE towards the outer regions. The maps 
B and C are obtained with larger area bins and we can see that the details disappear in these plots. On the
whole we find that the central regions have an average age of the LSFE as $\sim$ 20 Myr, whereas the periphery has an average age of about 80 Myr. This is similar to what was found in the MCPS maps. Since the OGLE III maps
have the highest spatial resolution, we tried to identify any propagating star formation in the bar
region. We do not detect any propagation along the bar. All along the bar, from the north-west end to the
south-east end, we detect pockets of very young stars, as young as $\le$ 10 Myr. These pockets are surrounded by slightly older stars ($\sim$ 20 Myr),
as seen in map A. The medium resolution map, B also shows that young stars are distributed right across the
bar suggesting the star formation has been active all along the bar in small pockets. The map A has missing
regions towards the periphery which means that these regions have age of the LSFE older than 120 Myr. Similar to figure 6, the maps B and C cover more of the outer regions, which is due to larger bin area.

The maps obtained from both the data sets suggest that the inner regions have continued to form stars up to
$\le$ 10 Myr, whereas the outer regions stopped forming stars earlier ($\sim$ 80 Myr). This would suggest an inward quenching of star formation. The outer regions are older with an age of about 80 Myr suggesting that the star formation stopped at around this age. The younger ages for the LSFE for the inner regions suggest that the star
formation has stopped relatively recently or still continuing to form stars. The younger LSFE regions located in the inner regions have
a clumpy distribution which suggests that the star formation has broken up into smaller pockets. Around these
pockets, one can see a gradation in the age with relatively older stars located in the periphery.
To summarise, the age maps suggest an outside to inside quenching of star formation in the inner three
degrees of the LMC, in the age range 80 - 1 Myr, with the inner regions experiencing star
formation till very recently.  Even though three maps correspond to three different area bins and hence the estimated age of the LSFE differ slightly, above result is seen in all the maps, with varying details. 

The statistical distribution of the age of the LSFE in the LMC is shown in figure 8. The histograms in three  colours represents distribution of ages derived with three different area bins as shown in the figure. In both the distributions using OGLE III \& MCPS (upper \& lower panel in the figure), it is seen that as we go to higher and higher spatial resolution, the peaks of the distribution tends to shift towards older ages. This can be explained in terms of the number of stars present in a subregion of a particular area. As we make finer bins, the number of stars in the MS of the CMD becomes less and less, which will lead to an older MS turn-off. In the case of OGLE III, the age of the LSFE peaks at 60-70 Myr and 30 - 40 Myr, for small area bins. In the case of large area bins, the peak is found only at 30 - 40 Myr. The time at which the largest number of regions stopped forming stars (30 - 40 Myr) is similar for all the three area bins. Since OGLE III scans do not cover the northern star forming regions, the above distribution
is applicable to the central regions, including the bar. Thus, we might conclude that most of the central regions stopped forming stars at about 30 - 40 Myr. All the area bins also show an isolated peak at 90-100 Myr.
This peak is also found in the MCPS distribution for all the area bins. This peak may be similar to the 100 Myr star formation peak identified by H\&Z09.

In the case of the MCPS data, we see that the smallest area bin shows a peak at 30 - 40 Myr, whereas the largest area bin shows a peak at 0 - 10 Myr. The progressive shifting of the peak to younger ages with the increase in the area binned can be clearly seen. In fact, the medium resolution map
shows that the peak to be between 0 - 40 Myr. Thus, MCPS data which has a larger area coverage finds that the star
formation stopped at 0 - 10 Myr for most of the regions. Thus, the influence due to the northern regions is to make the peak shift to younger ages, suggesting that the star formation in the northern regions continued to younger ages, when compared to the central regions. Thus, we find a peak of star formation at 0 - 10 Myr, similar to the 12 Myrs star formation peak found by H\&Z09, for the MCPS data. The
peaks of star formation identified here coincides  with those found by H\&Z09, even though the methods
used are different. The above fact also suggests that the ages derived by this method are comparable with
those derived by them. 

Among the maps presented above, the OGLE III maps help to understand the finer details of star formation in the central
regions, due to the higher spatial resolution (the smallest area for subregions). The MCPS maps cover a larger area, in particular, the northern regions. It can be seen that there are a number of star forming regions in the north of the LMC, but there are very less in the southern LMC. With respect to the
 optical center at
RA = 5$^h$19$^m$38$^s$; Dec = -69$^{\circ}$27"5.2' (J2000.0 \cite{df73}), one can notice a lopsidedness in the recent star formation towards the northern regions. That is, the quenching of star formation is not symmetric with respect
to the center of the LMC. The quenching appears to be more effective in the southern LMC, when compared to the northern part. In order to substantiate this point, we plot the LSFE age map (the high resolution map (A) in figure 6) in the xy plane using the optical center to convert RA-Dec to x - y plane (figure 9). We also show concentric circles of radii 2, 3 and 4$^o$  with respect to the center. It can be seen that in the southern regions, the older ages appear in the inner circles, between radii 2$^o$ and 3$^o$. In the northern regions, older ages appear only in the outermost annulus and beyond 4$^o$. This map clearly shows the lopsidedness and the extension of the younger star forming regions to the north and north-east. The map also suggests that the southern regions are more or less symmetric with respect to the center. This result is consistent with the presence of star forming regions like LMC1, LMC5, LMC4 and the super giant shells in the north, and 30 Doradus in the north-east, whereas similar regions are not found in the southern LMC. The map presented above is in the sky plane and it needs to be deprojected on to the LMC plane, in order to comment on the lopsidedness. That is, in order to understand these features, one needs to study their location in the plane of the LMC and not in the sky plane. This is presented in the following section.

\subsubsection{Deprojection of data to the LMC plane}
In order to study the lopsidedness of the age distribution, we need to deproject the data to obtain the distribution in the plane of the LMC. The LMC plane is inclined with respect to the sky plane by an angle $i$ (with face on view corresponds to i = 0) and the PA of the line of nodes (measured counterclockwise from north) is $\Theta$. The near side of the LMC plane lies at $\Theta_{near}=\Theta$-90 and the far side $\Theta_{far} = \Theta$+90. Conversion of RA$-$Dec to Cartesian co-ordinates x,y (Van der Marel $\&$ Cioni 2001) is done using the following conversion equation,\\
\begin {center}
x($\alpha$,$\delta$) = $\rho$ cos($\phi$)\\
y($\alpha$,$\delta$) = $\rho$ sin($\phi$)
\end {center}
where $\rho$ $\&$ $\phi$ are the angular coordinates of a point defined by the coordinates ($\alpha$, $\delta$) in the celestial sphere, where $\rho$ is the angular distance between the points($\alpha$, $\delta$) and ($\alpha_0$, $\delta_0$) which is defined to be the center of the LMC, and $\phi$ is the position angle of the point ($\alpha$, $\delta$) with respect to ($\alpha_0$, $\delta_0$). By convention,  $\phi$ is measured counter clockwise starting from the axis that runs in the direction of decreasing RA, at constant declination $\delta_0$. Correction for the PA and $i$ can be applied if we know the mean distance to the LMC center $D_0$, using conversion equations,
\begin{equation} 
x' =  \frac {D_0 cos (\phi-\Theta_{far}) sin \rho cos i }{ cos i cos \rho - sin i sin \rho sin (\phi - \Theta_{far})}
\end{equation}
\begin{equation}
y' = \frac {D_0 sin (\phi - \Theta_{far}) sin \rho }{ cos i cos \rho - sin i sin \rho sin (\phi - \Theta_{far})}
\end{equation}
But, in practice it is useful not to use the coordinates in the LMC disk plane, but a new system, in which the line of nodes lie at the same angle in the (x", y") plane of the LMC, as in the projected (x,y) plane of the sky. It is obtained by rotating (x',y') by an angle $\Theta_{far}$,
\begin{equation}
x" = x' cos\Theta_{far} - y' sin\Theta_{far}
\end{equation}
\begin{equation}
y" = x' sin\Theta_{far} + y' cos\Theta_{far}
\end{equation}

%% Figure 10 (MCPS - deprojected with rings)
\begin{figure}
%\\begin{minipage}{152mm}
   \resizebox{\hsize}{!}{\includegraphics{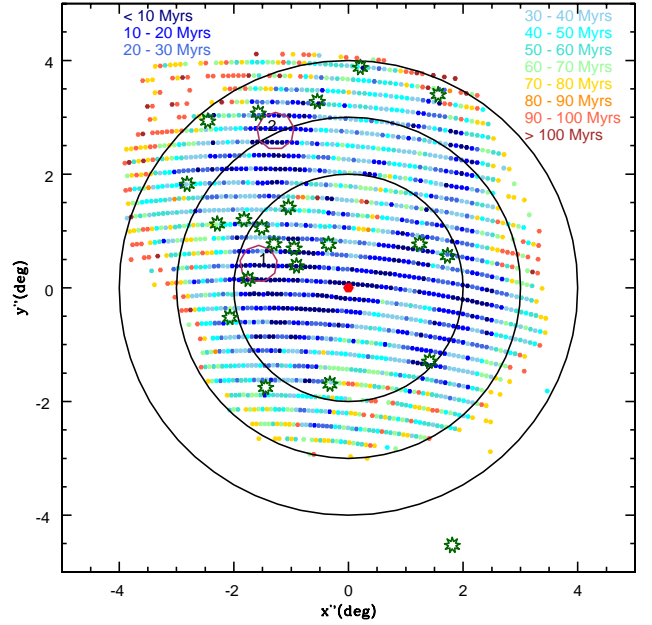}}
   \caption{The LSFE map of the LMC similar to figure 9, in the deprojected x" - y" plane of the LMC, concentric circles are drawn centered at optical center with radii as in fig 9.  The relevant features identified in the LMC plane are shown as  hexagons. The numbering is decoded as 1. 30 Doradus, 2. Constellation III(\cite{m80}), The dark green points are the HI  super giant shells (\cite{k99})} 
%3. NorthWest void, 4. Blue arm, 5. SouthEast arm, 6. NorthWest arm.(H\&Z09)}}
    %\\end{minipage}
    \end{figure}

The deprojected MCPS map is shown in figure 10, where concentric circles of radii 2$^o$, 3$^o$ and 4$^o$ are also shown to compare the distribution with figure 9. This map also shows the lopsidedness, suggesting an extension in the north and north-eastern directions. If we consider regions which stopped star formation around 40 Myr or younger, the location of such regions gives an impression that it is being stretched in the north east-south west direction, with more
such regions located in the north-east, with respect to the LMC center. Thus, in the plane of the LMC, the recent star formation has a lopsidedness towards the north and north-east, which was suggestive in the earlier maps. The direction in which the distribution appears to be stretched is in the direction of our Galaxy. \cite{vc01} found an elongation in the outer stellar distribution of the LMC disk, when viewed in the LMC plane. This elongation in the recent star forming regions is also in the similar direction. We compare this lopsidedness with respect to the distribution of HI gas in the following section.

%% Figure 11 (HI clouds)
%% Figure 12 (clusters)
\begin{figure}
%\\begin{minipage}{152mm}
   \resizebox{\hsize}{!}{\includegraphics{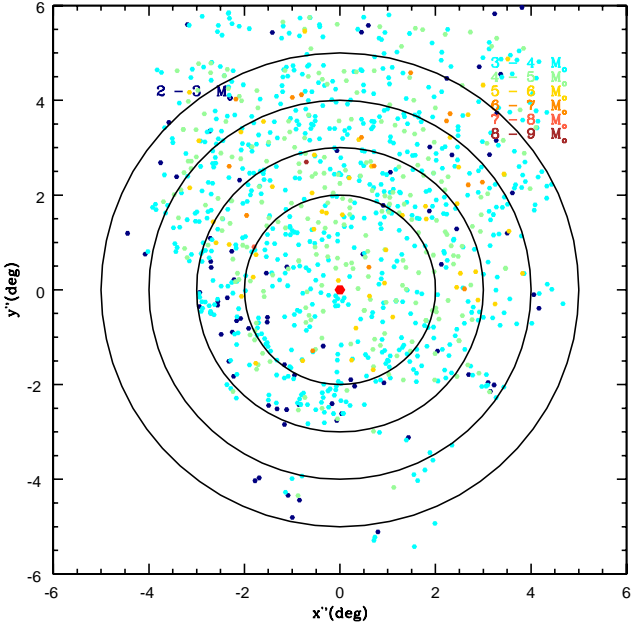}}
	\caption{Map of HI clouds in the LMC plotted in x" - y" plane. Colour coding is according to the mass in log scale, as specified in the figure. Data is from \cite{k07}. The LMC optical center is shown as a red point. Concentric rings are 
over plotted at radii 2$^o$, 3$^o$, 4$^o$ \& 5$^o$ .}
%\\end{minipage}
    \end{figure}
\begin{figure}
%\\begin{minipage}{152mm}
   \resizebox{\hsize}{!}{\includegraphics{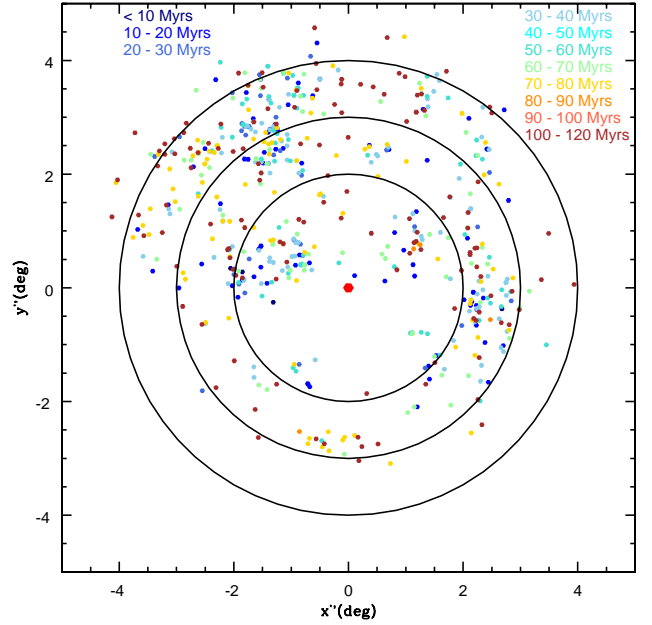}}
   \caption{
The spatial distribution of young clusters ($<$ 120 Myr)
from \cite{g10} is plotted in the deprojected plane of the LMC. Color coding is according
to cluster age as specified in the figure. The red dot is the optical center of the LMC. Concentric 
rings are drawn at radii 2$^o$, 3$^o$ \& 4$^o$.}
    %\\end{minipage}
    \end{figure}
\subsubsection{Comparison with HI clouds and star clusters}
The results obtained above suggest that the recent star formation in the LMC has a lopsidedness in the north and north-east direction. Then, one would also expect to see a similar lopsidedness in the distribution of the HI clouds {\it in the LMC plane}. We have
plotted the HI clouds using the data from \cite{k07} in the LMC plane in figure 11. The colour code
used is according to the mass of the cloud, which is indicated in the figure. Concentric circles at 2$^o$, 3$^o$, 4$^o$ and 5$^o$ are also shown. The deprojected HI distribution is seen to be lopsided with respect to the center. Most of the clouds are located to the north of the center, with very less clouds in the south. The clouds are located within 3$^o$ in the south, while, they are distributed beyond 5$^o$ in the north.
%In order to see the distribution of the massive clouds, HI clouds with mass greater than  10000$M_{\odot}$ are only used. The size of the points in the figure is proportional to the mass the of the cloud.%
 The massive HI clouds are also preferentially populated in the north compared to the south. Thus the distribution of the HI clouds show lopsidedness towards the north. The age maps as well as the HI distribution correlate well and point to an extension of the LMC disk towards the north with respect to the optical center. On the other hand, the recent star formation also shows a north-east extension. This might suggest that the star formation is more efficient in the north as well as the north-east in converting gas to stars. 

The spatial distribution of young clusters (age $<$120 Myr) in the LMC plane is shown in figure 12. The optical center is shown along with concentric rings at radii 2$^o$, 3$^o$ and 4$^o$. The cluster distribution
is also found to be lopsided. The clusters in the age range 60 - 100 Myr are distributed up to 3$^o$ in the south, whereas they can be seen up to 4$^o$ and slightly beyond in the north. The clusters in the above age range show lopsidedness towards north. The distribution of clusters younger than 40 Myr is found to shrink to inner regions and are concentrated at the ends of the bar and in the northern regions. We can also notice a large concentration of clusters in the north-east.  Thus the young ($\le$ 40 Myr) clusters seem to accumulate in the north east-north direction, except for a small group in the south-west of the bar, probably due to the presence of the bar. We summarise that clusters in the age range 60 - 100 Myr show lopsidedness towards north, whereas clusters younger than 40 Myr show lopsidedness towards north and north-east. This might suggest that
the north-east enhancement in the star formation is likely to have happened in the last 40 Myr, whereas the northern enhancement is seen in the last 100 Myr.

In order to compare the ages of LSFE estimated here with the ages of clusters, we have plotted clusters
in three age groups in figure 13. The LSFE age map only shows the age of the last star formation event and does not suggest anything about previous star formation episodes in the region. Therefore, while
comparing with the cluster ages, we expect that the ages of the youngest clusters in a given region should match with the LSFE ages. In the figure, clusters younger than 40 Myr are shown in the top-left panel. We see a good correlation between the locations of clusters and sub regions in this age range such that clusters are
located near sub regions with similar ages. This also suggests that the ages estimated for the 
sub-regions are 
similar to the ages of youngest clusters in the vicinity. We also notice that there are some regions in the bar which have been forming stars to very young ages, but there are no cluster formation in these regions. The plot on the top-right panel shows clusters in the range 40 - 70 Myr, and these clusters are distributed similar to the
younger clusters. We do not see any correlation as most of these clusters are located in region which continued to form stars and younger star clusters. In a few regions in the north-west , east and south, we see similar ages for cluster and the nearby sub-regions. The bottom-left panel shows the distribution of clusters in the 
age range 70 - 120 Myr. These clusters are distributed in a comparatively larger radius, than the young clusters. The location of these clusters match with the ages of the sub-regions in the outer regions of the map. Thus, we find that the age distribution of the LSFE  match well with distribution of the clusters.
The cluster distribution also suggests the shrinking of star forming regions to smaller pockets in the inner LMC, in the last 100 Myr.
%% Figure 13 (comparison map cluster vs LSFE) 
\begin{figure}
%\\begin{minipage}{152mm}
  \resizebox{\hsize}{!}{\includegraphics{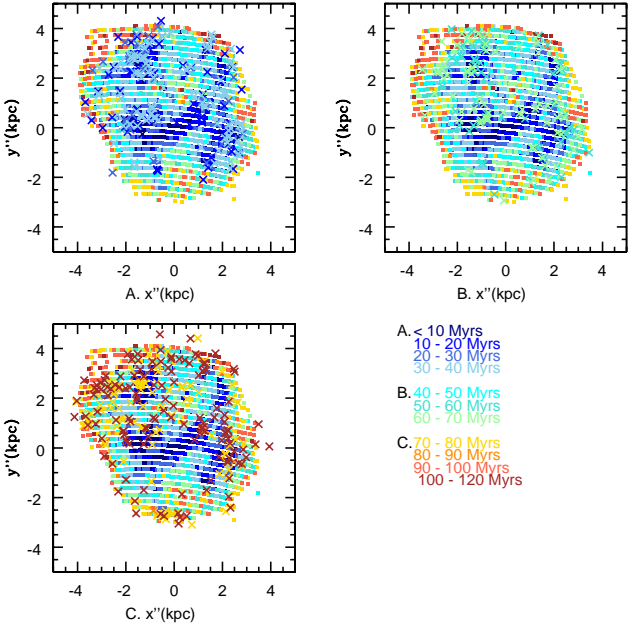}}
   \caption{The age distribution of young clusters (from \cite{g10}) are over plotted on the LSFE map of the LMC in the deprojected x" - y". Three different age groups are shown, top left, $\leq$ 40 Myr, top right, 40 - 70 Myr \& 
bottom left, 70 - 120 Myr. Color coding is according to the age as specified in the figure.} 
    %\\end{minipage}
   \end{figure}

%% Figure 13 (comparison map) - may be we need to put this map
%\begin{figure}
%%\\begin{minipage}{152mm}
%  \resizebox{\hsize}{!}{\includegraphics{fig13.eps}}
%   \caption{Comparing with the star formation history from H\&Z09}
%    %\\end{minipage}
%   \end{figure}

\subsection{Shift in the center of young stellar distribution}
The LSFE age maps suggest that the young star forming regions are found to be lopsided to the north and north-east. The distribution of young clusters as shown in figure 12 suggested that the clusters younger than about 40 Myr show a preferential location to the north and north-east. The HI distribution suggests that most of the gas is located in the northern LMC disk. Thus, the center of the recent star formation seems to be shifted to north or north-east, with respect to the optical center. In order to study this shift in the distribution of young population as a function of age, we used the MCPS data. Since the number of clusters in a given age range is small, the center estimation is not done with clusters. 

We used MS star counts to estimate the distribution of stars younger than a particular age using the MCPS data. In the MS, stars brighter than a cut-off magnitude are identified. These stars are younger than the age corresponding to the cut-off magnitude. The center of the distribution of these stars are estimated in the sky plane (RA vs DEC plane) as well as the LMC plane (x" vs y" plane). The age tagged with such a population will be the age of the oldest population in the group and it will have stars younger than this age. Table 1 contains the age of the oldest population of the group, centers in RA and Dec, and x" and y", the number of stars considered for the center estimation and the error in the values of the center. x" \& y" are in kpc where 1$^o$ is equal to 0.89 kpc at the distance of the LMC. The centers of stellar population in the LMC plane, for various ages are shown in figure 14. The optical center (OC) and the kinematic center of the gas (KC, taken from \cite{k98}) are also shown. The oldest population considered is about 500 Myr and the youngest is about 10 Myr.  The center of the distribution does not shift between 500 - 200 Myr, even though a small shift towards the south can be noticed. The shift between 200 - 40 Myr is clearly visible along the y" axis, suggesting a northward shift. We detect a shift of 7 pc/10 Myr towards north between 200 - 100 Myr, while an enhanced shift of 27 pc/10 Myr towards north is detected in the 100 - 40 Myr age range. On the other hand, no significant shift is present along the x" axis (east-west axis) in the above age range. For population younger than 40 Myr, a shift in both the axes can be noticed suggesting that the center is progressively shifting in the north-east direction. In the 40 - 10 Myr age range, a shift of 50 pc/10 Myrs to the north \& 28 pc/10 Myrs to the east are detected. This analysis suggests that the northern lopsidedness in the stellar distribution started between 200 - 100 Myr. This can be compared to the
appearance of the northern blue arm in the age range 160 - 100 Myr in the SFH by H\&Z09, which could shift the center of the stellar distribution to the north. To summarise, we find that the center of the distribution of stars shifts northward in the age range 200-40 Myr, and the center is found to shift in the north-east direction for population younger than 40 Myr. This correlates well with the age vs shift found in the cluster distribution. Thus the stellar population as well as the cluster population has experienced a shift in the north-east direction only in the last 40 Myr. H\&Z09 also finds an enhanced star formation in the north-eastern regions for ages $<$ 50 Myr, which can be inferred from their figure 8.

\begin{table*}
      \caption[]{The centers of the stellar population in the LMC for various ages, using MCPS data.}
         \label{Table:1}
	\centering	
	\begin{tabular}{c | c | c | c | c | c | c | c}
	\hline
	Age (Myr) & RA(deg) & Dec (deg) & $x^"$ (kpc) & $\sigma$$x^"$ & $y^"$ (kpc) & $\sigma$$y^"$ & N\# \\ 
	\hline
	   8 & 80.5303 & -68.5728 & -0.2598 & -0.0137 & 0.8000 & 0.0160 & 13327   \\
	  18 & 80.4294 & -68.6190 & -0.2218 & -0.0080 & 0.7584 & 0.0094 & 36536  \\
	  40 & 80.3217 & -68.7290 & -0.1756 & -0.0046 & 0.6505 & 0.0052 & 107243  \\
	  60 & 80.3163 & -68.8027 & -0.1666 & -0.0034 & 0.5780 & 0.0038 & 192619  \\
	  93 & 80.3297 & -68.8768 & -0.1632 & -0.0025 & 0.5051 & 0.0027 & 353207 \\
	 214 & 80.3542 & -68.9608 & -0.1589 & -0.0014 & 0.4184 & 0.0015 & 1141153 \\
	 493 & 80.3146 & -68.9201 & -0.1451 & -0.0009 & 0.4470 & 0.0010 & 3151134 \\
	\hline
\end{tabular}
   \end{table*}

%%Figure 14 (centroids deprojected LMC plane)
\begin{figure}
%\\begin{minipage}{152mm}
   \resizebox{\hsize}{!}{\includegraphics{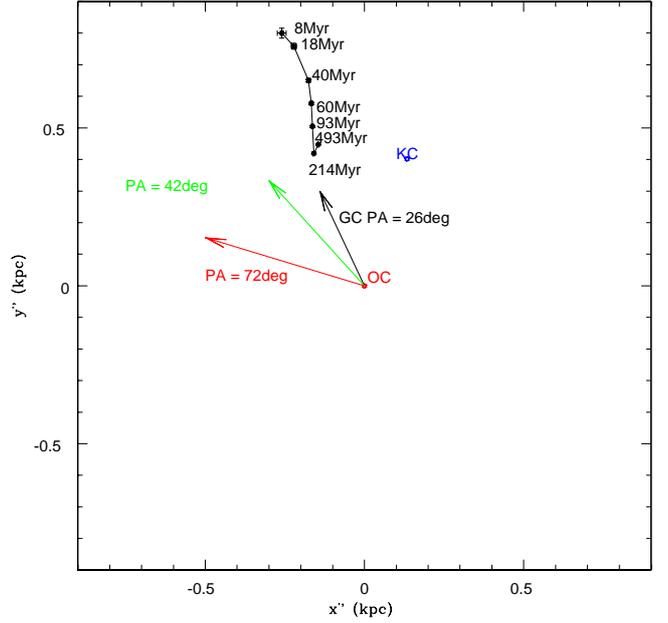}}
   \caption{The centers of the stellar population in the LMC for various ages, is shown in the plane of the LMC, with error bars. The direction of velocity vector of the LMC is shown in red at a position angle of 72$^o$, (calculated from the proper motion values provided by \cite{p08}) \& the line of interaction of the MW \& the LMC according to our convention at a position angle of 26$^o$. The direction given in \cite{v01} is shown in green. KC is the HI kinematic center (\cite {k98}) shown in blue, \& OC is the optical center of the LMC
shown as a red point.}  
%the degeneracy of points shown here, with L\&SMC are
%the centroids of iso density contours in each of the clouds}
    %\\end{minipage}
    \end{figure}
 The line of interaction between the MW \& the LMC is shown in figure 14. According to our convention, this line  is at a position angle of 26$^o$ (shown in black), whereas the direction given in \cite{v01} is 42$^o$ (shown in green). This is the direction to the Galactic center. The direction of velocity vector of the LMC is shown in red, at a position angle of 72$^o$.  It can be seen that the direction of shift of the center is almost in the direction towards the Galactic center. The LMC disk is known to be inclined such that the north-east part is closer to the Galaxy. The lopsidedness in star formation is in the same direction of the inclination. Also, the LMC is moving past our Galaxy after the closest approach. 
Thus the lopsidedness in the stellar as well as HI distribution to the north may be due to  the gravitational attraction of our Galaxy on the gas of the LMC disk and the enhanced compression in the northern regions. The movement of the LMC could cause compression of the gas in the north- eastern side resulting in enhanced star formation in the north-east. The center shifts and the time-scales derived in this section can be used to understand the details of the above two processes on the gas resulting in star formation.
 We shall discuss these aspects in detail in the section for discussion.

\subsubsection{Comparison with the star formation history of \cite{hz09}}
A complete star formation history of the LMC was derived by H\&Z09 using the multi band photometry of the MCPS. They provided the star formation rate (SFR) in $M_{\odot}/Myr$, for particular age bins for different regions in the LMC. Our analysis identifies the LSFE in a region and estimates its age, which is quite different from the analysis of H\&Z09. Hence, it is not possible to compare the two results quantitatively. On the other hand, we compare our results with the recent SFH as estimated by H\&Z09. We compare our figure 9 with the 12.5 Myr and 6.3 panels of figure 8 in H\&Z09. It can be seen that there is a very good correlation between the star forming regions identified in figure 8 and those in our figure 9. The shrinking of star formation to smaller regions is clearly seen if we compare their 12.5 Myr and 6.3 Myr panels. They find that the northern blue arm appears in the 160-100 Myr age range. We notice a shift in the population to the north in the last 200 Myr. Also, the enhanced star formation in the north-east regions appear for ages less than 50 Myr in H\&Z09, whereas we find such an enhancement at 40 Myr. They identified  peaks of star formation at 12.5 Myr and 100 Myr, and this correlates well with the peaks of 0 - 10 Myr and 90 - 100 Myr identified by us. Thus, the ages of the LSFE estimated here correlates well with the recent SFH derived by H\&Z09.
% We have simulated synthetic CMDs for both the clouds to quantify the statistical effects which will be discussed in detail in the Discussion. We also estimate the Star Formation Rate (SFR) threshold required for our analysis for various age ranges. An age map of youngest population which satisfy the corresponding SFR cut-off is estimated from H\&Z09 shown in Figure 15. The area binning followed by H\&Z09 is 12 x 12 arcmin$^2$, and 24 x 24arcmin$^2$. %Our LSFE maps with all the three area sampling  are also shown for comparison, same as Figure.6. 
%The ages younger than 100 Myr, from H\&Z09 are 6.3, 12.5, 25, \& 50 Myr. In a spatial comparison with our LSFE map, Figure 6 B or C, we could compare the dark blue, blue, cyan, brownish red points only. Since H\&Z09 doesn't have 60 to 90 Myr, we can't directly compare the yellow, orange, red points. In this view point we can see the maps resembles pretty well, with youngest population 6.3 to 25 Myr, shrunk to the central region (blue points) and there are older regions 50, 100 Myr, in the periphery (cyan \& brownish red points).

%%figure (H&Z spatial map)
%\begin{figure}
%%\\begin{minipage}{156mm}
%   \resizebox{\hsize}{!}{\includegraphics{new.eps}}
 %  \caption{The age map of youngest population from H\&Z09. Color coding is according to the age as shown in the color bar. }
%    %\\end{minipage}
%\end{figure}
%% figure 15 (SMC extinction - MCPS map)
\begin{figure}
%\\begin{minipage}{156mm}
   \resizebox{\hsize}{!}{\includegraphics{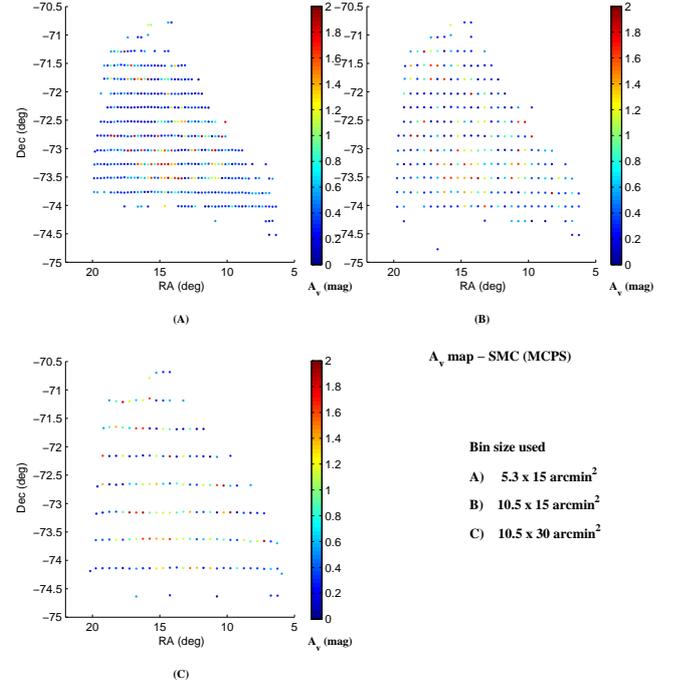}}
   \caption{The Extinction map of the SMC using MCPS data in the RA-Dec plane for three different area binning, as specified in the figure. Color coding is according to the $A_v$ value, which varies from 0.2 to 2.0 as shown in the color bar.}
    %\\end{minipage}
    \end{figure}
%% figure 16 (SMC extinction - OGLEIII map)
\begin{figure}
%\\begin{minipage}{156mm}
   \resizebox{\hsize}{!}{\includegraphics{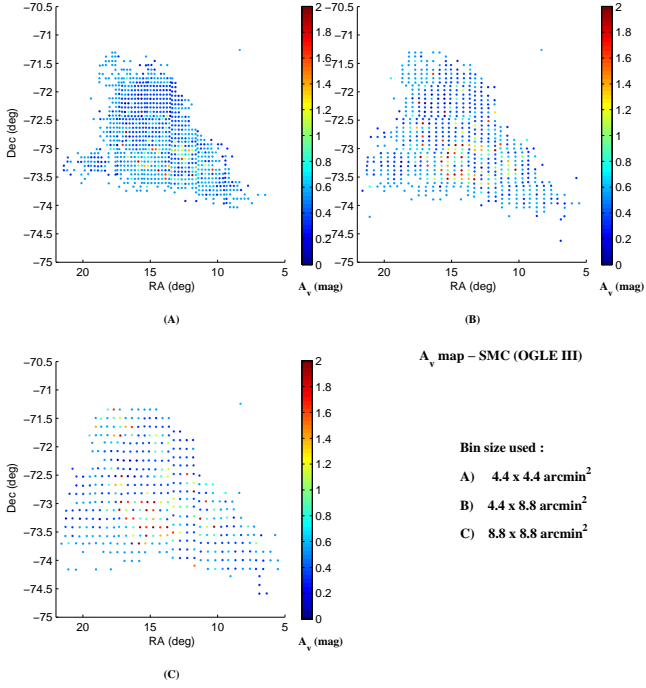}}
   \caption{The SMC extinction map for three area bins, similar to figure 15, using OGLE III data.}
    %\\end{minipage}
    \end{figure}
%% figure 17 (SMC extinction - comparison with Zaritsky)
\begin{figure}
%\\begin{minipage}{152mm}
   \resizebox{\hsize}{!}{\includegraphics{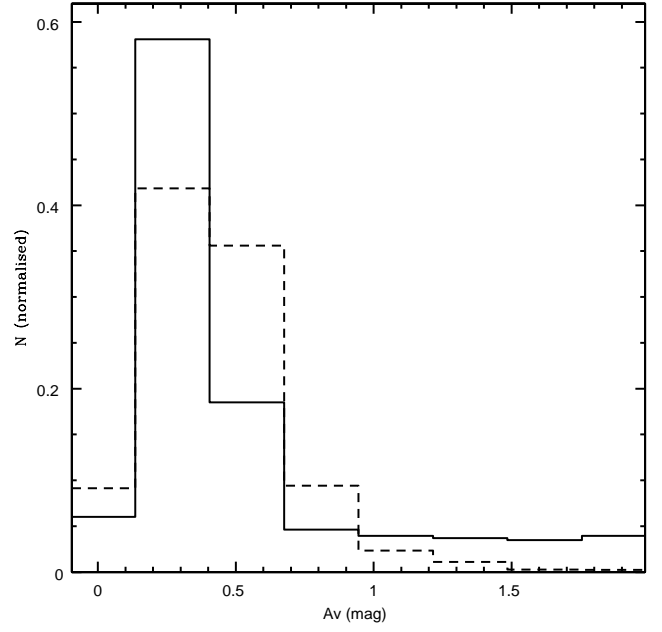}}
   \caption{The estimated distribution of extinction
	    in the SMC (shown as solid line) is compared with  distribution obtained from the extinction map of hot stars provided by \cite{z02} (shown as dashed line, extracted from the fits image from the website).}
    %\\end{minipage}
    \end{figure}	

\subsection{SMC: Extinction} 
The extinction towards the SMC is estimated using the MCPS and OGLE III data sets. As in the case of the LMC, we estimated the reddening E(V$-$I) of the main sequence stars near the turn-off for each region.
%In the case of the SMC, the area bin used is relatively large and hence the number of points is less. 
The extinction ($A_V$) maps estimated using the MCPS data are shown in figure 15 and those estimated from OGLE III are shown in figure 16. The area used for three different bin sizes are mentioned in the figure. The maps are found to be similar with the extinction in the range, 0.2 - 2.0 mag. The average extinction is found to be between 0.2 - 0.5 mag with central regions showing a higher value of extinction. The eastern wing also has regions with large extinction. \cite{z02} have estimated the extinction across the SMC. We compare our estimation with their values in the form of distribution, in figure 17. In general, both distributions match very well and both have a peak in the range, 0.2 - 0.4 mag. Similar to the case of the LMC, we derive lesser extinction when compared to \cite{z02}, due to the reasons mentioned in section 4.1. We do observe a mild inconsistency in the distribution for extinction values higher than 1.5 magnitude. Since the number of regions involved is small, it is unlikely to affect the results derived below.

%% figure 18 (LFSE - MCPS)
\begin{figure}
%\\begin{minipage}{156mm}
   \resizebox{\hsize}{!}{\includegraphics{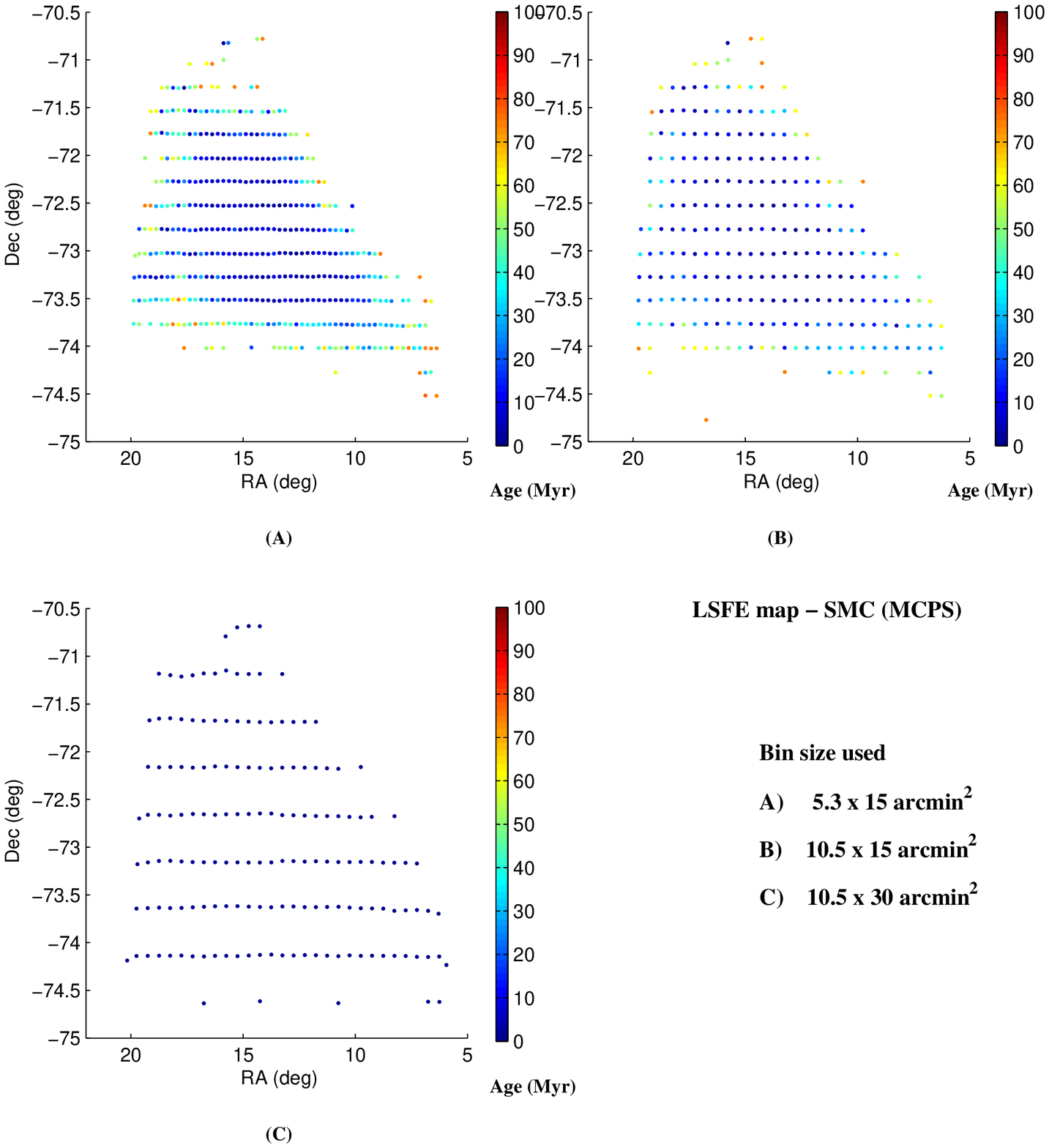}}
   \caption{The LSFE map of the SMC using MCPS data in the RA Dec plane with three different area binning, as specified in the figure. Color coding is according to the LSFE age as shown in the color bar.}
    %\\end{minipage}
    \end{figure}
%% figure 19 (LSFE - OGLEIII)
\begin{figure}
%\\begin{minipage}{156mm}
   \resizebox{\hsize}{!}{\includegraphics{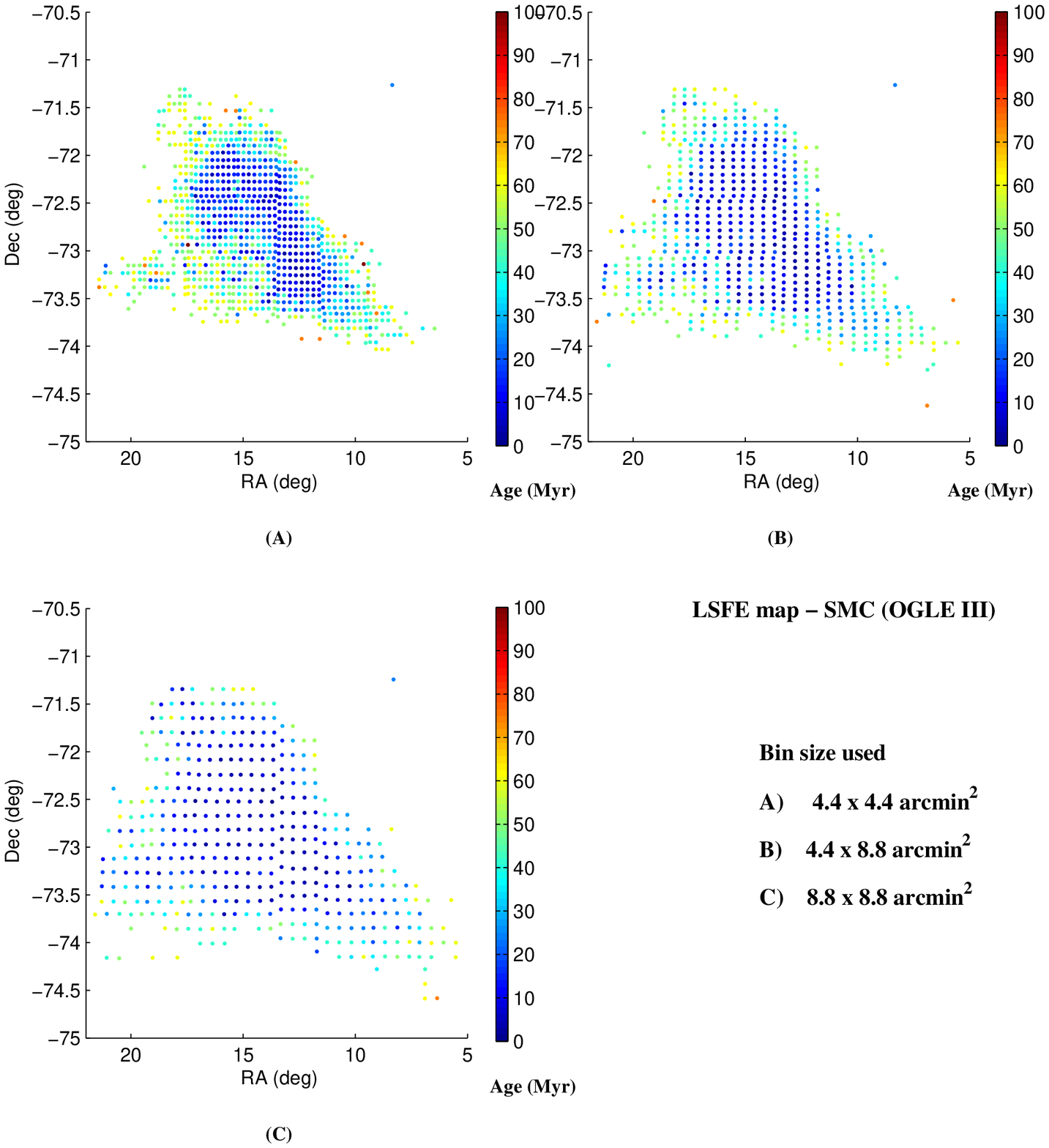}}
   \caption{The LSFE map of the SMC similar to figure 18, using OGLE III data}
    %\\end{minipage}
    \end{figure}
%% figure 20 ( SMC age histogram)
 \begin{figure}
%\\begin{minipage}{152mm}
   \resizebox{\hsize}{!}{\includegraphics{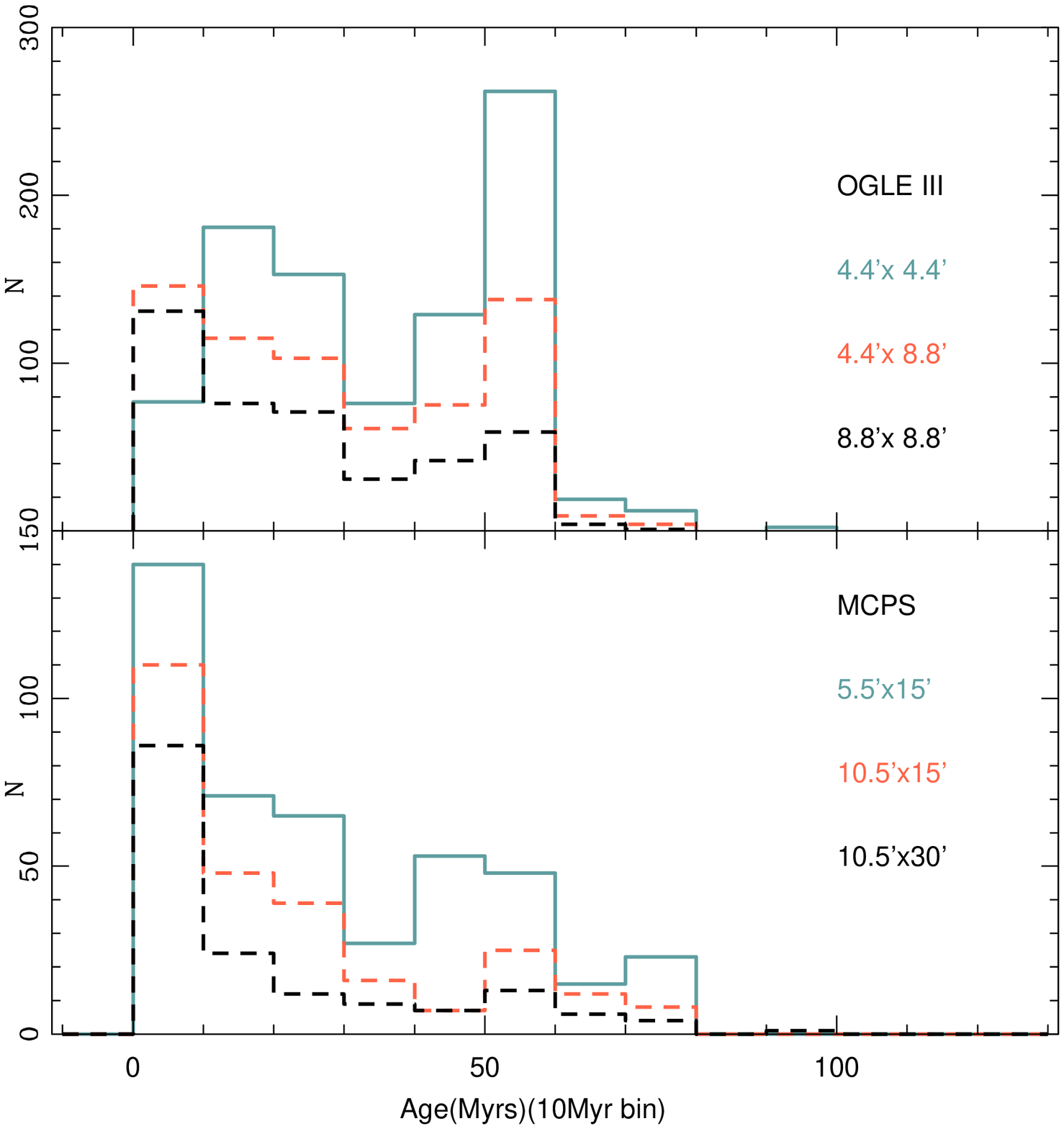}}
   \caption{The statistical distribution of LSFE ages for the SMC. Upper panel shows OGLE III data and 
lower panel shows MCPS, with the three colors corresponds to different area, as specified in the figure.}
    %\\end{minipage}
    \end{figure}
\subsection{SMC: age map of the LSFE}
The spatial distribution of age of the LSFE in RA vs Dec plane is estimated and the maps are presented in figures 18 (MCPS) and 19 (OGLE III). We have adopted three sizes of area (bin sizes as specified in the figure) to estimate the ages and to understand the effect of the area on the estimated age. The MCPS maps show that the central region has experienced star formation till very recently (0 - 20 Myr). In the high resolution (smallest area size) map A, the outer regions appear to have older ages of about 60 - 80 Myr. This is not so clear in the map B, and in map C, we do not see any such increase in the age of the LSFE with radius. This clearly shows the effect of area used in estimating the age of the LSFE.
The OGLE III plots use smaller sizes of area (figure 19), and the high resolution map A, shows that the inner region has a substructure with two clumps in the north-east and south-west direction. This substructure is similar in location to the two HI super shells, 304A and 37A (\cite{st99}). The map also suggests that the north-east wing has more regions with recent star formation, when compared to the southern clump, which is located near the center of the SMC. This structure disappears in the maps B and C. In all the OGLE III maps, we can find that the age of the LSFE in the central regions is in the age range 0 - 20 Myr, while the periphery shows an age of $\sim$ 60 Myr. Unlike in the LMC, we do not see a gradation in age, increasing outward from the center, even though the outer regions show older age. We can also notice that the missing regions due to the limiting magnitude appear only in the periphery, suggesting that all the inner regions have experienced star formation in the 100 Myr. The missing outer regions also suggest that the star formation stopped in the outer regions much earlier. With the increase in the area of the bin (maps B and C), more regions get added to the periphery due to the shift of the LSFE to younger ages.
To summarise, we find that the star formation in the inner SMC is not very structured, when compared to the LMC. We could identify the eastern wing in the map. Most of the inner regions have experienced star formation in the last 0 - 20 Myr. Similar to the LMC, we find a marginal evidence for outside to inside quenching of star formation in the last 60 Myr. 

The histogram of the age distribution of the LSFE is shown in figure 20. All the three maps from the OGLE III (upper panel) identify a peak at 50 - 60 Myr. The younger peak shifts from 10 - 30 Myr in the highest resolution to 0 - 10 Myr in the low resolution maps. The shift to younger ages is noticed with the increase of the area considered.
The distributions obtained from the MCPS map (lower panel) show a mild peak at 50 - 60 Myr in the lowest two resolutions, whereas the peak widens to 40-60 Myr for high resolution. All the three resolutions show a peak at 0 - 10 Myr. In summary, we find that most of the regions in the SMC have experienced star formation till very recently, with one peak at 0 - 10 Myr and another peak at 50 - 60 Myr. The SFRs of the SMC are shown in figure 19 of H\&Z09, which is derived from the the SFH estimated in \cite{hz04}. They identified a peak around 50 Myr and another around 10 Myr, which is in good agreement with our finding. Hence the results derived here correlate well with the recent SFH derived by \cite{hz04}. 

\subsubsection{Comparison with HI distribution \& Star clusters}
%%figure 21 & 22 (HI distribution in SMC, xy plane, LSFE map, xy plane)
\begin{figure}
%\\begin{minipage}{152mm}
   \resizebox{\hsize}{!}{\includegraphics{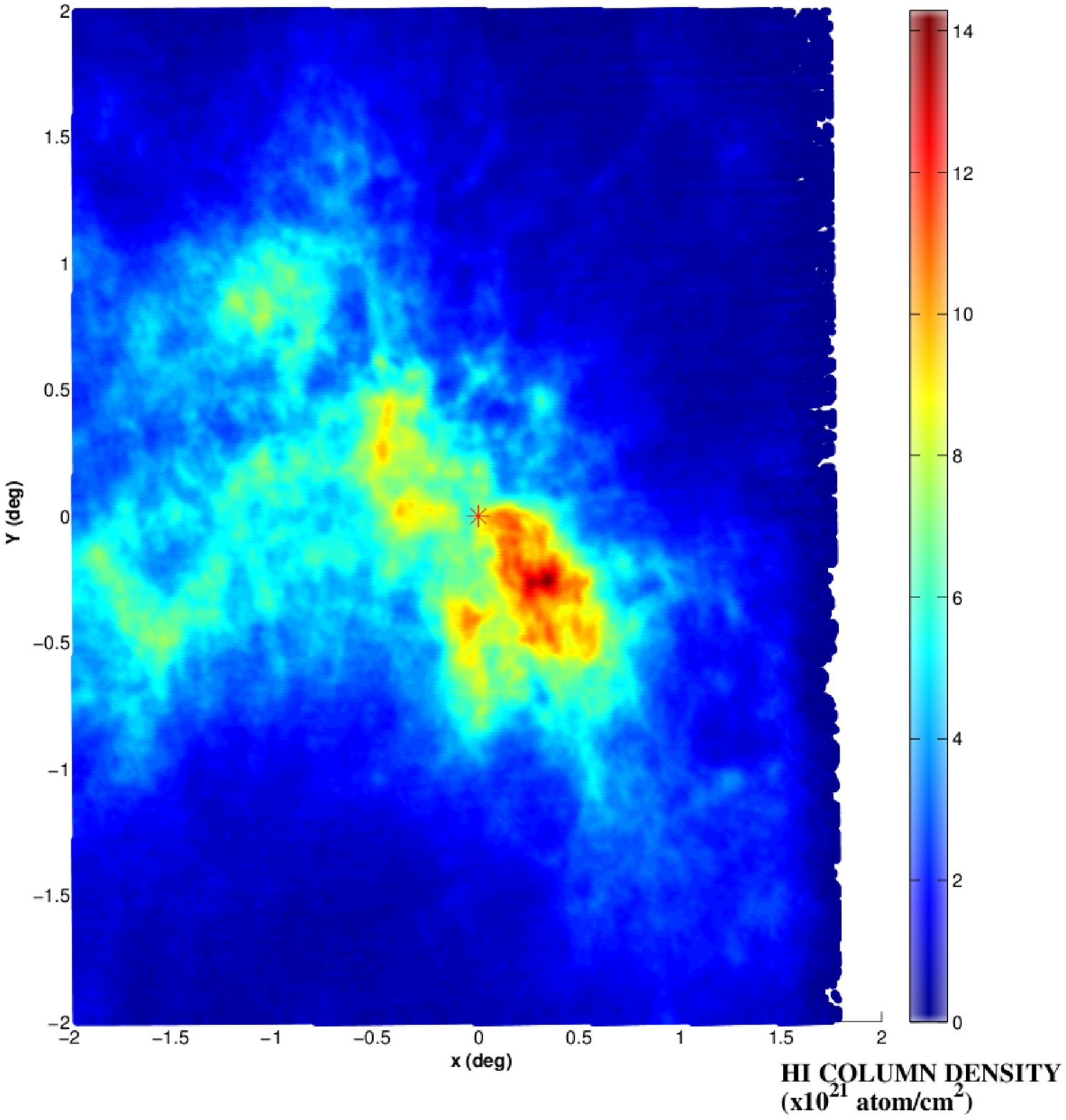}}
   \caption{The HI column density distribution in the SMC is shown in the projected x - y plane of sky. Color coding is according to HI column density as depicted in the color bar. Data is taken from \cite {st04}.}  
    %\\end{minipage}
    \end{figure}

\begin{figure}
%\\begin{minipage}{152mm}
   \resizebox{\hsize}{!}{\includegraphics{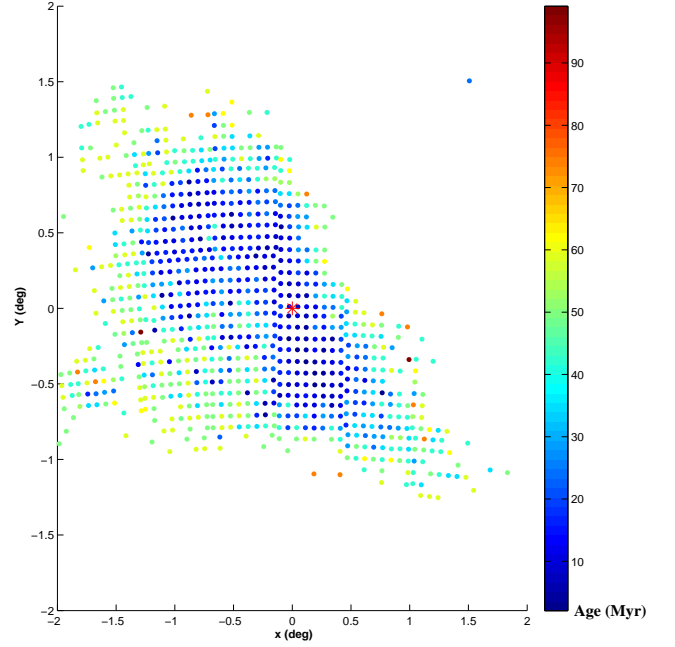}}
   \caption{The LSFE map of the SMC is shown in the projected x - y plane of the sky, similar to figure 19 A. The optical center is shown in red.}  
    %\\end{minipage}
    \end{figure}
%figure 23 SMC clusters
\begin{figure}
%\\begin{minipage}{152mm}
   \resizebox{\hsize}{!}{\includegraphics{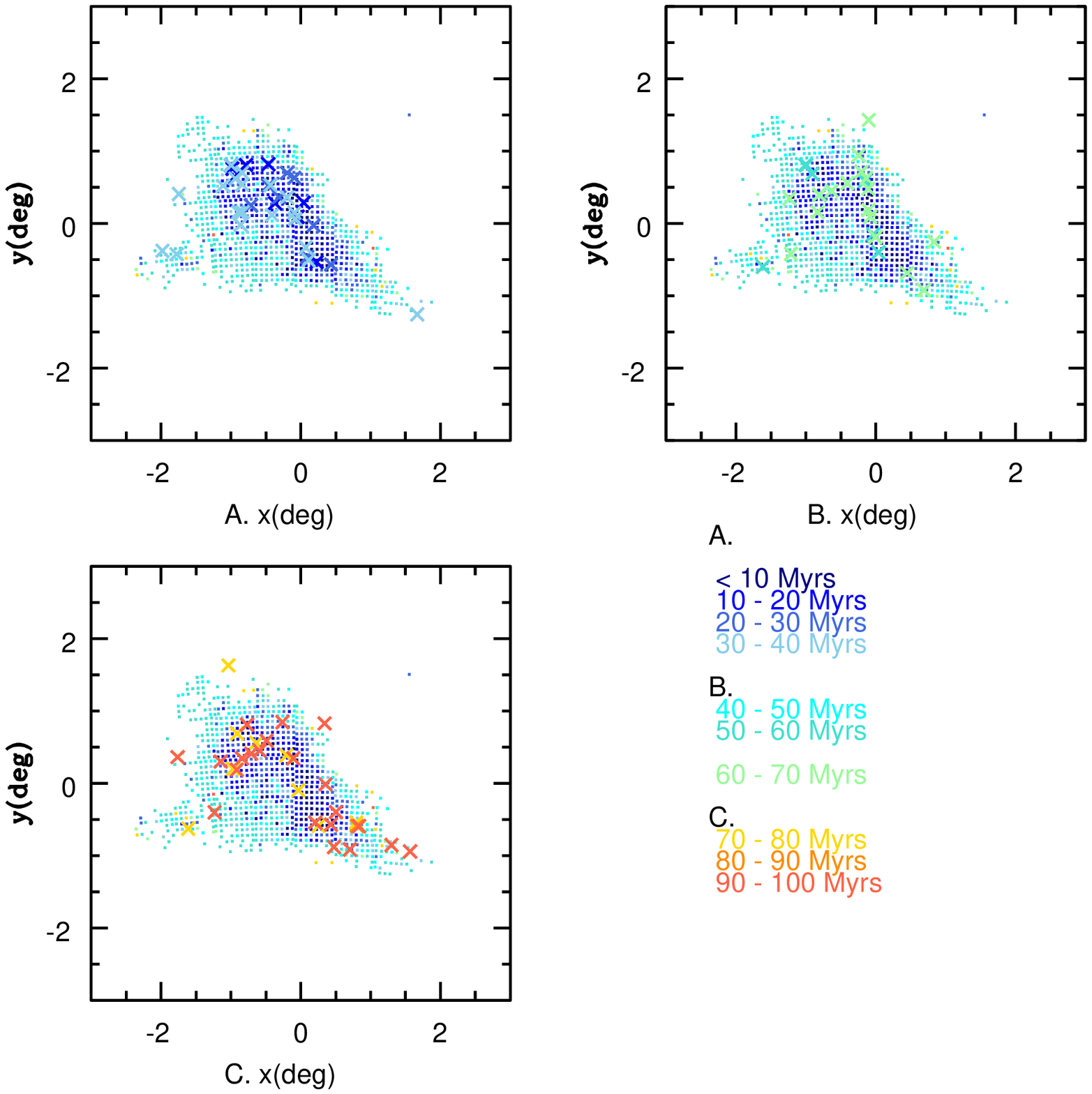}}
   \caption{The age distribution of clusters as old as or younger than 100 Myrs (\cite{g10}) over plotted with the LSFE map (same as figure 22)  in the projected x - y plane of the sky. Three different age groups are shown, top left $\leq$ 40 Myr, top right 40 - 70 Myr \& 
bottom left 70 - 100 Myr. Color coding is according to the age as specified in the figure.}  
    %\\end{minipage}
   \end{figure}

We have plotted the HI column density distribution of \cite {st04} in figure 21. The HI has the
largest column density in the south-western part. The LSFE age map (corresponding to the high resolution map A in figure 19) is shown in the x - y plane in figure 22 for comparison. The north-east and the south-west substructures of the LSFE age map match well with the location of the HI. The LSFE map also identifies young star forming regions in the eastern part of the wing as well as in the north-east part. Overall, the young star forming regions identified here coincide with the locations of HI with column density more than $7-8 \times 10^{21} atom/cm^2$. We identify more regions with recent star formation in the north-east substructure when compared to the south-west substructure, whereas relatively more/dense HI is found in the south-west substructure. This might suggest that the star formation is more efficient in converting gas to stars in the north-eastern wing. This might also suggest an efficient gas compression in the north-east, similar to the case of the LMC.

The ages of young star clusters given in \cite{g10} is over plotted on the LSFE maps in figure 23. As mentioned in section 4.2.2, while
comparing with the cluster ages, we expect that the ages of the youngest clusters in a given region should match with the LSFE ages. The top-left panel shows clusters in the age range 0 - 40 Myr. These clusters are found to be located near regions with recent star formation, in general. Clusters in the age range 30 - 40 located towards the outer regions, are also found near sub-regions with similar ages. Young clusters (0 - 30 Myr) are found to be located in the north-east and south-west substructures and relatively
more number of clusters are found in the north-east substructure. This suggests that the cluster formation is more efficient in the north-east substructure. The top-right panel shows clusters in the age range 40 - 70 Myr. We notice older clusters near younger star forming regions suggesting older cluster forming episodes in these regions. Clusters in the age range 70 - 100 Myr are shown in the bottom-left panel. Since we do not see many regions with LSFE in this age range, comparison is not possible. The older clusters are found to occupy a relatively large radial extent compared to the younger clusters. This mildly suggests the outside to inside quenching of star/cluster formation. 

\subsection{Shift in the center of the young stellar distribution in SMC}
We used MS stars younger than various age cut-offs to study the shift of centroids in the SMC. The data used here is OGLE III, taking in to account its higher resolution and larger number of stars. Since the geometry of the SMC is not well understood, the center of the distribution of these stars are estimated in RA and Dec and in the x \& y coordinates (with respect to the optical center of the SMC). Optical center is taken as RA = 0$^h$52$^m$12.5$^s$; Dec = -72$^{\circ}$49"43' (J2000.0 \cite{df73})
 x \& y are in kiloparsec where 1$^o$ corresponds to 1.04 kpc at the distance of the SMC.  Table 2 contains the age of the oldest population of the group, centers in RA \& Dec, and x \& y, the number of stars considered for the center estimation and the error in the values of the center. The oldest population considered is about 450 Myr and the youngest is about 5 Myr. The location of the centers, along with the ages and error bars are shown in figure 24. We detect a shift in the center between 500 - 200 Myr in the north-east direction. We estimate a shift of 2.1 pc/10 Myr along the x and 2.2 pc/10 Myr along the y directions. The shift then continues upto 30 Myr in the same direction, with an enhanced rate of 4 pc/10 Myr along the x direction and 5 pc/10 Myr along the y direction in the 200 - 30 Myr age range. There is no significant center shift for ages younger than 30 Myr. We see a mild shift to the west between 14-6 Myr, but is within 2$\sigma$ of the error, as can be seen from figure 24. The amount of shift in the center is much less compared to the shift we detect in the LMC. Thus, we detect a significant shift in the center of the population younger than 500 Myr in the north-east direction. This is the direction towards the LMC. H\&Z09 detected a coincident peak of enhanced star formation at 400 Myr in both the Clouds and they suggested that this may be due to their mutual interaction. The above center shift between 500 - 200 Myr may be due to this star formation episode and enhanced star formation in the north-eastern region. This may also be due to the appearance of the wing in this age range. We detect an enhanced center shift in the 200 - 30 Myr age range, this could be caused by  the gravitational attraction of our Galaxy during the intergalactic passage.  

\begin{table*}
      \caption[]{The centers of the stellar populations in the SMC for various ages using OGLE III data.}
         \label{Table:2}
	\centering	
	\begin{tabular}{c | c | c | c | c | c | c | c}
	\hline
	Age (Myr) & RA(deg) & Dec (deg) & $x$ (kpc) & $\sigma$$x$ & $y$ (kpc) & $\sigma$$y$ & N\#  \\ 
	\hline
	   6 & 14.2007 & -72.7345 & -0.3631 & -0.0197 & 0.0445 & 0.0128 & 2430 \\
	  14 & 14.3156 & -72.7533 & -0.3981 & -0.0109 & 0.0472 & 0.0069 & 7968 \\
	  33 & 14.2957 & -72.7728 & -0.3916 & -0.0063 & 0.0334 & 0.0041 & 23377 \\
	  79 & 14.1303 & -72.8157 & -0.3413 & -0.0036 & -0.0088 & -0.0024 & 67919  \\
	 188 & 14.0868 & -72.8496 & -0.3279 & -0.0021 & -0.0425 & -0.0014 & 203383 \\
	 445 & 13.9135 & -72.9043 & -0.2739 & -0.0011 & -0.0980 & -0.0008 & 597714 \\
	\hline
\end{tabular}
   \end{table*}
%%%%%%%%

%%%%%figure 24 - plot the center shift in the SMC 
\begin{figure}
%\\begin{minipage}{152mm}
   \resizebox{\hsize}{!}{\includegraphics{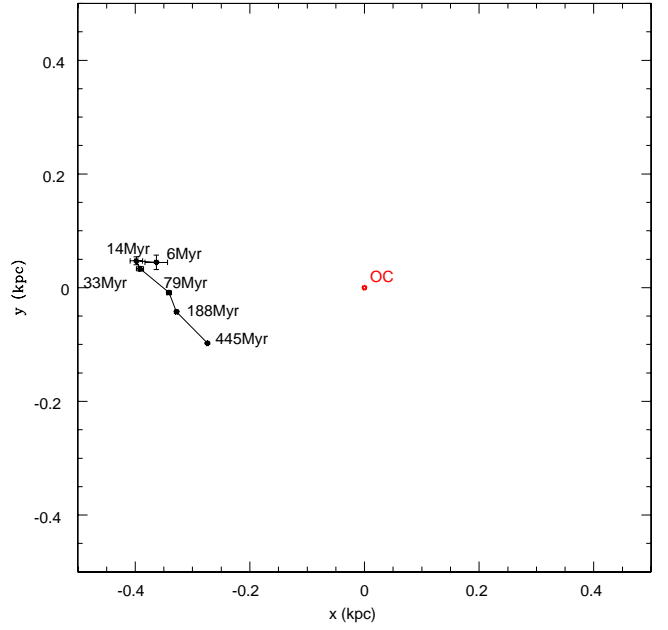}}
   \caption{The figure depicts locations of center of number density distribution of stars tagged with various upper age cut-off in the projected sky plane for the SMC (table 2). The error bars are shown \& the optical center is shown in red.}  
    %\\end{minipage}
    \end{figure}
\section{Discussion}
Recent high-precision proper motion estimates suggest that the LMC and the SMC are either on their first passage or on an eccentric long period ($>$ 6 Gyr) orbit about the MW. This differs markedly from the canonical picture in which the Clouds travel on a quasi-periodic orbit about the MW (period of ~2 Gyr). Without a short-period orbit about the MW, the origin of the Magellanic Stream, a young (1-2 Gyr old) coherent stream of H I gas that trails the Clouds, spans 150$^o$ across the sky, can no longer be attributed to stripping by MW tides and/or ram pressure stripping by MW halo gas (taken from Besla et al. 2010). Also, the episodes of star formation
found in the LMC has been traditionally attributed to its repeated interaction with the MW and the SMC. SMC being a smaller galaxy, it is unlikely to create large impact on the LMC. In the first passage scenario, the LMC has probably had its closest approach at about 200 Myr ago. This event is likely to have caused a significant impact on both the Clouds, especially on the LMC. 

In the plane of the LMC, we find that the present distribution of the HI gas is lopsided towards the north. 
The star formation in the last 40 Myr is lopsided towards north and north-east. The center of the stellar population is found to shift to north in the last 200 - 40 Myr, and towards north-east in the last 40 Myr. Similarly, H\&Z09 found the appearance of the blue arm around 160 -100 Myr ago and the north-eastern enhancement for ages $<$ 50 Myr.  In order to understand the MW-LMC-SMC interaction, their locations are shown in the plane of the LMC, in figure 25. The direction of velocity vector of the LMC is at a position angle 72$^o$ (shown in red). The location of the galactic center is shown according to our convention (PA = 26$^o$)and (\cite{v01}) (PA=42$^o$, shown in green). The direction of motion of the LMC is in the same quadrant as the line connecting the LMC and the GC, which is the north-east quadrant. The location of the SMC is such that the LMC lies in the line of interaction of the SMC \& the MW. This line is also in the north-east direction. It is quite possible that the recent star formation in the Clouds is due to the effect of the gravitational attraction of the MW on the HI gas in them.

 We propose the following scenario to explain the lopsidedness in the LMC. We suggest that the HI gas in the LMC is shifted/pulled to the north due to the perigalactic passage at about 200 Myr. The gas in the north is shocked/compressed to form stars starting from about 200 Myr, due to the motion of the LMC in the MW halo. The efficiency in converting gas to stars seems to have increased in the 100 - 40 Myr age range. The direction of the center shift is more or less in the direction of the line connecting the LMC and the Milky.  At about 40 Myr, we detect the north-eastern regions to have enhanced star formation suggesting an efficient compression/shocking in the north-east. The LMC is moving away from the MW, after the closest approach.
This could result in the north-east to be compressed than the north, which is the direction of motion of the LMC. We do not detect any age difference or propagating star formation along the northern blue arm. The north-east enhancement in star formation is probably due to the compression
of gas due to the motion of the LMC. A similar scenario of bow-shock induced star formation was proposed by \cite{dbr98}. They suggested that the gas gets compressed in the east and then moves towards north with the LMC disk rotation. This would suggest a gradation in the age along the northern region from the east. Such a gradient is not seen here. We suggest that this compression has been active only in the last 40 Myr and not before that. 

Another result of this study is the outside to inside quenching of star formation within the last 100 Myr. We identify a peak in the star formation at 90 - 100 Myr, which is also identified by H\&Z09. The peak of star formation found at 100 Myr is probably a global feature in the LMC, which is probably the effect of the perigalactic passage on the HI gas of the LMC. This star formation was present in most of the regions and
 the star formation then became restricted to the inner LMC. Thus the star formation was quenched from outside to inside after the 100 Myr peak. In the central regions, there is an indication that most of the regions stopped forming stars in 30 - 40 Myr age range with only a few pockets continuing to form stars. On the other hand, the north-east regions, south-west end of the bar and some northern regions continue to form stars till very recently, giving rise to the 0-10 Myr peak in star formation. The quenching of star formation was efficient in the southern LMC as most of the gas present was converted to stars and there is not much gas left. The star formation in the north and the north-east continues due to the presence of gas and due to efficient compression produced by the motion of the LMC.

The case of the SMC is found to be a bit different. We detect peaks of star formation at 0-10 Myr and at 50 - 60 Myr. The recent star formation is not found to be as structured as in the LMC. We detect a shift in the center of the stellar population younger than 500 Myr. This might be due to the interaction of the SMC with the LMC at about 400 Myr, as suggested by the coincident star formation (H\&Z09). We detect an enhanced shift in the center of the population in the 200 - 40 Myr age range, which is similar to that found in the LMC. The directions of the lines connecting the LMC and the Galaxy to the SMC, as shown in figure 25, are similar. This means that the effect due to the LMC and the Galaxy on the SMC will be in the similar direction. Thus, shift in the center
of the stars in the 200 - 40 Myr age range may be a combined effect of the LMC as well as the perigalactic passage. In the SMC, we also see that the north-eastern regions are active in star formation, when compared to the south and the west. This may also be due to the effective compression of the HI gas in the north-eastern regions of the SMC, due to the motion of the LMC-SMC system in the halo of the MW.

In summary, the recent star formation in the LMC has been dictated by the last perigalactic passage.
The timescales and locations of star formation identified in this study are valuable to model the recent interactions between the Clouds and the MW. The lopsidedness of the HI distribution towards the north can give constraints on the parameters governing the gravitational force of the Galaxy on the LMC, before, during and after the perigalactic passage. The compression of gas in the northern regions during the perigalactic passage and the compression of gas in the last 40 Myr also can give constraints on the direction of motion of the LMC as well as the effect of the Galactic halo on the HI gas. It will be interesting to bring out these details in the LMC disk, using an SPH simulation. The recent star formation in the SMC is complicated because of the combined gravitational effect of the LMC and the Galaxy, especially with both of them located in the same direction. Thus, disentangling the effect on the SMC due to its interaction with the LMC and the effect due to perigalactic passage may be difficult.

\begin{table*}
      \caption[]{The input parameters to the synthetic CMD and the estimated LSFE, $A_v$ values in the L\&SMC. The extinction applied for synthesising CMD are 0.55 for the LMC \& 0.46 for the SMC.}
         \label{Table:3}
	\begin{center}	
	\begin{tabular}{c | c | c | c | c | c }
	\hline
	
	& log(age) &  nstar\# & Mass($M_\odot$) & LSFE log(age) & derived $A_v$ \\ 
	\hline

LMC &  7.40   &    500   &   1124.5   &   7.88  &  0.7018                    \\        
   & &   1000   &   2361.3   &   7.90  & 0.4538                             \\
   & &  1500   &   3670.9   &   7.49  &  0.5878                             \\
   & &  2000   &   4980.1   &   7.34  &  0.5878                   \\
   & &  2500   &   6258.8   &   7.41  &  0.5878                            \\
   & &  3000   &   7546.2   &   7.41  &  0.5878                              \\
   & &  3500   &   8824.9   &   7.41  &  0.5878                              \\
   & &  4000   &  10133.7   &   7.41  &  0.5878                              \\
 SMC & 7.50   &    500   &   1123.1   &   8.11  &  0.4290                              \\
     &   &   1000   &   2372.7   &   7.95  &  0.4712                              \\
     &   &   1500   &   3666.4   &   7.62  &  0.5332                          \\
     &   &   2000   &   4987.3   &   7.54  &  0.5630                              \\
     &   &   2500   &   6299.9   &   7.54  &  0.5630                              \\
     &   &   3000   &   7663.7   &   7.54  &  0.5630                              \\
     &   &   3500   &   8990.9   &   7.54  &  0.5630                              \\
     &   &   4000   &  10392.2   &   7.54  &  0.5630                              \\ 
   
	\hline
\end{tabular}
\end{center}
   \end{table*}

\begin{table*}
      \caption[]{The error in the derived ages estimated using synthesised data. The tabulated error holds for synthetic
CMDs with minimum mass shown in column 4.}
         \label{Table:4}
	\centering	
	\begin{tabular}{c |c | c | c }
	\hline
	& log(age) & error in log(age) & mass (M$_\odot$) \\ 
	\hline
	LMC & 7.0  & 0.15   & 7550   \\
	    & 7.10 & 0.15   &  4980     \\
	    & 7.35 & 0.10   &  4274     \\
	    & 7.40 & 0.09   &  3671      \\
	    & 7.50 & 0.07   &  4986    \\
	    & 7.60 & 0.06   &  3668    \\
	    & 7.70 & 0.06   &  4280    \\
	    & 7.75 & 0.04   &  3657   \\
	    & 7.90 & 0.02   &  2367  \\
	    &  8.00 & 0.03  & 2362   \\
       \hline
       SMC & 7.0 & 0.15  & 6298 \\
	   & 7.2 & 0.17  & 6310 \\
	   & 7.3 & 0.15  & 4975 \\
	   & 7.4 & 0.14  & 3680 \\
	   & 7.5  & 0.12  & 3666  \\
	   & 7.55 & 0.11  & 3669  \\
           & 7.65 & 0.12  & 3680  \\
	   & 7.70 & 0.07  & 3669  \\
	   & 7.80 & 0.07  & 2356  \\
	   & 7.90 & 0.05  & 2375  \\
	& 8.00 & 0.02  & 2366  \\
	\hline
\end{tabular}
   \end{table*}

%%figure 26 (location of LMC, MW and SMC)
\begin{figure}
%\\begin{minipage}{152mm}
   \resizebox{\hsize}{!}{\includegraphics{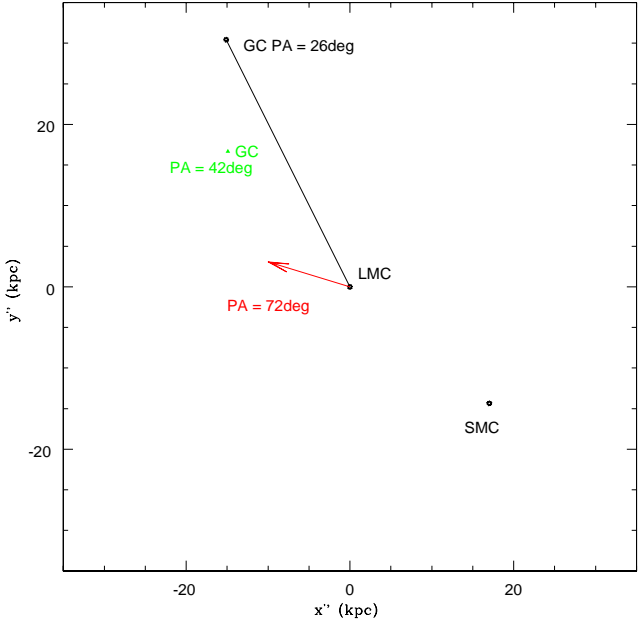}}
   \caption{The locations of the MW, LMC, \& the SMC are shown in the plane of the LMC, the direction of velocity vector of the LMC is shown in red at a position angle 72$^o$, \& the line of interaction of the MW \& the LMC according to our convention is drawn at a position angle of 26$^o$. The Galactic Center taken from \cite{v01} is shown in green at a PA 42$^o$. The location of the SMC is such that the LMC lies in the line of interaction of the SMC \& the MW.}  
%the degeneracy of points shown here, with L\&SMC are
%the centroids of iso density contours in each of the clouds}
    %\\end{minipage}
    \end{figure}
\section{Appendix: Error analysis}
In this study, we have used a simple technique of identifying the MS turn off and converting into the age of the youngest star formation. The ages estimated by this method, in general, are found to be comparable with the cluster ages and ages estimated by H\&Z09. The ages estimated here are affected by the photometric errors, errors in the estimation of extinction, and the finite resolution in binning along the luminosity as well as the colour axes. We have derived the error in the estimated values of extinction and LSFE age using two methods, (1) synthetic CMD and (2) propagation of errors. These are described below.

 We synthesised CMDs with an IMF 2.35, in the mass range 0.6 - 20 M$\odot$, incorporating the observed (typical) photometric error and using the  \cite{m08} isochrones for  metallicities 0.008 (LMC) \& 0.004 (SMC). Synthetic CMDs are created for turn off ages between 10 to 100 Myr. In order to quantify the sampling effects and errors on the derived turn off age, we performed the same analysis on the synthesised data. For each population, we estimated the LSFE age and extinction. In order to understand the effect of sampling, the analysis 
is repeated by varying the total number of stars in a CMD. We applied an $A_v$ of 0.55 mag for the LMC \& 0.46 mag for the SMC to synthesise the CMDs. The LSFE age \& extinction estimated from the synthetic CMDs are compared with the input values  to find the error. In order to incorporate the effect of sampling, number of stars (or total mass) is also varied. Synthetic CMDs are created with a wide range in the number of stars (more as well as less than that found typically in the observed CMDs).  As an example, the input parameters for the synthetic CMD and the estimated results for the L\&SMC for a single age are tabulated in table 3. Columns 1, 2, \& 3 are the log (age), the total number of stars \& total stellar mass in the synthetic CMD. The total stellar mass can be used to estimate the typical star formation rate which is detected using this method (and the cut-off) as a function of age. The star formation rate can be estimated by dividing the  total mass with the age range corresponding to the $M_v$ and the bin size. Column 4, 5 are the estimated LSFE log (age) \& extinction. The analysis is done for various age ranges and the results are tabulated in Table 4. The columns are (log) age,  estimated error in age and the minimum mass (and hence the number of stars) required to produce the synthetic CMD, with the tabulated parameter. This number is found to be between 1500 -2000 and is smaller than the number of stars found in the observed CMDs above the limiting magnitude. The maximum error in the extinction is found to be within 2.48 times the bin size (0.1 mag) of the (V$-$I) colour. The error in log (age) is found to be in the range of 0.01 - 0.15 in the case of the LMC and 0.01 - 0.17 in the case of the SMC. This error includes the error in the estimation of extinction and
magnitude of the MS turn-off, statistical error and a typical photometric error. This method gives an estimate of error as a function of the LSFE age.\\

The error in the estimated LSFE as a function of location is estimated using the second method. The error is calculated by the propagation of error method, starting with the photometric error in V \& I bands, $\sigma$V \& $\sigma$(V-I).\\ 
$\sigma A_v$ = $ 2.48  \sqrt {\sigma{(V-I)}^2 + {(V-I)}_{bin}^2}$ \\
$\sigma$$M_v$ = $ \sqrt {\sigma V^2 + V_{bin}^2 + \sigma A_v^2}$ \\
$\sigma$age = constant x $\sigma$$M_v$ \\
$V_{bin}$ \& ${(V-I)}_{bin}$ are half the bin sizes used for magnitude \& color binning,  $\sigma$$M_v$ is the error in absolute magnitude $\sigma$$A_v$ is error in the estimated extinction \& $\sigma$age is the error in LSFE age in a logarithmic scale. The spatial plots of the errors are shown in figures 26 to 29, for L\&SMC for both the data sets MCPS \& OGLE III. The error in log(age) varies from 0.06 to 0.12. The error is found to be higher along the bar of the LMC \& the central regions of SMC. This is primarily the effect of photometric errors due to the crowding, as these are the densest regions in the respective galaxies. We identify larger error for the OGLE III data when compared to the MCPS data for the central regions of the SMC. The error in the LSFE age, as estimated by the above two methods are comparable. The first method is likely to estimate the true error and the second method is likely to over estimate the error. Both the methods
are included to show the variation of error as a function of age and location.
%%Figure 26,27,28,29 (Error maps)

\begin{figure}
%\\begin{minipage}{156mm}
   \resizebox{\hsize}{!}{\includegraphics{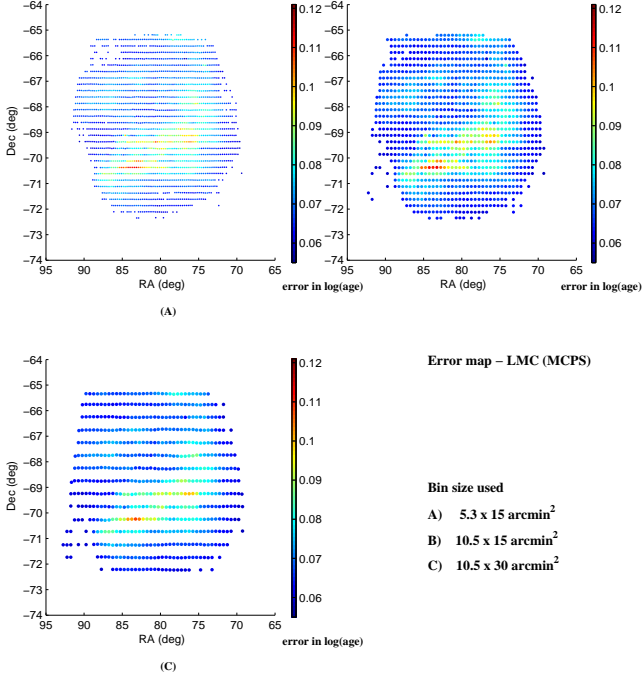}}
   \caption{The error in estimated age for the LMC using MCPS data for all the three area bins as indicated in the figure. Color coding is according to the error in log(age).} 
%\\end{minipage}
    \end{figure}

\begin{figure}
%\\begin{minipage}{156mm}
   \resizebox{\hsize}{!}{\includegraphics{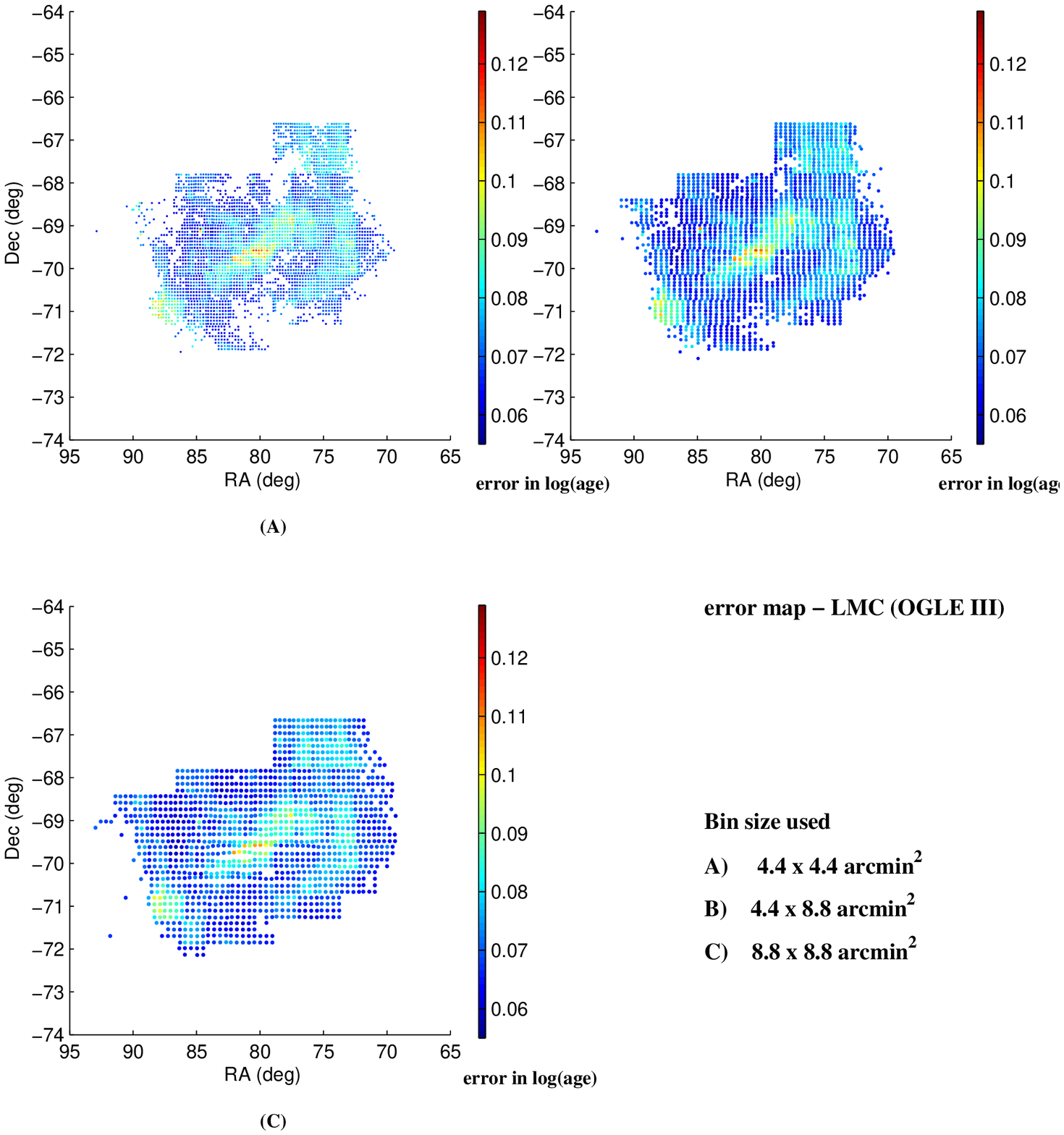}}
   \caption{Error map for the LMC using OGLE III, similar to figure 26. } 
%\\end{minipage}
    \end{figure}

\begin{figure}
%\\begin{minipage}{156mm}
   \resizebox{\hsize}{!}{\includegraphics{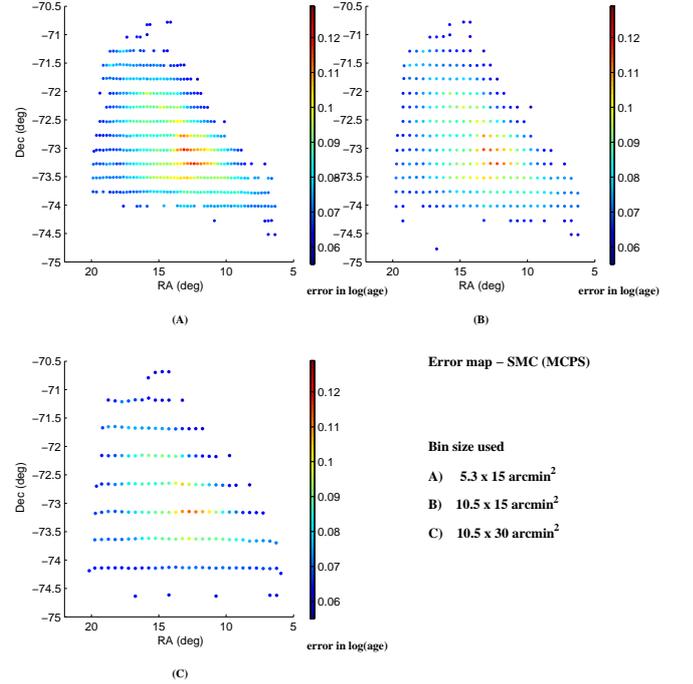}}
   \caption{Error in age estimation shown as a map for the SMC using MCPS data for all the three area bins as indicated in the figure. Color coding is according to the error in log(age).} 
%\\end{minipage}
    \end{figure}

\begin{figure}
%\\begin{minipage}{156mm}
   \resizebox{\hsize}{!}{\includegraphics{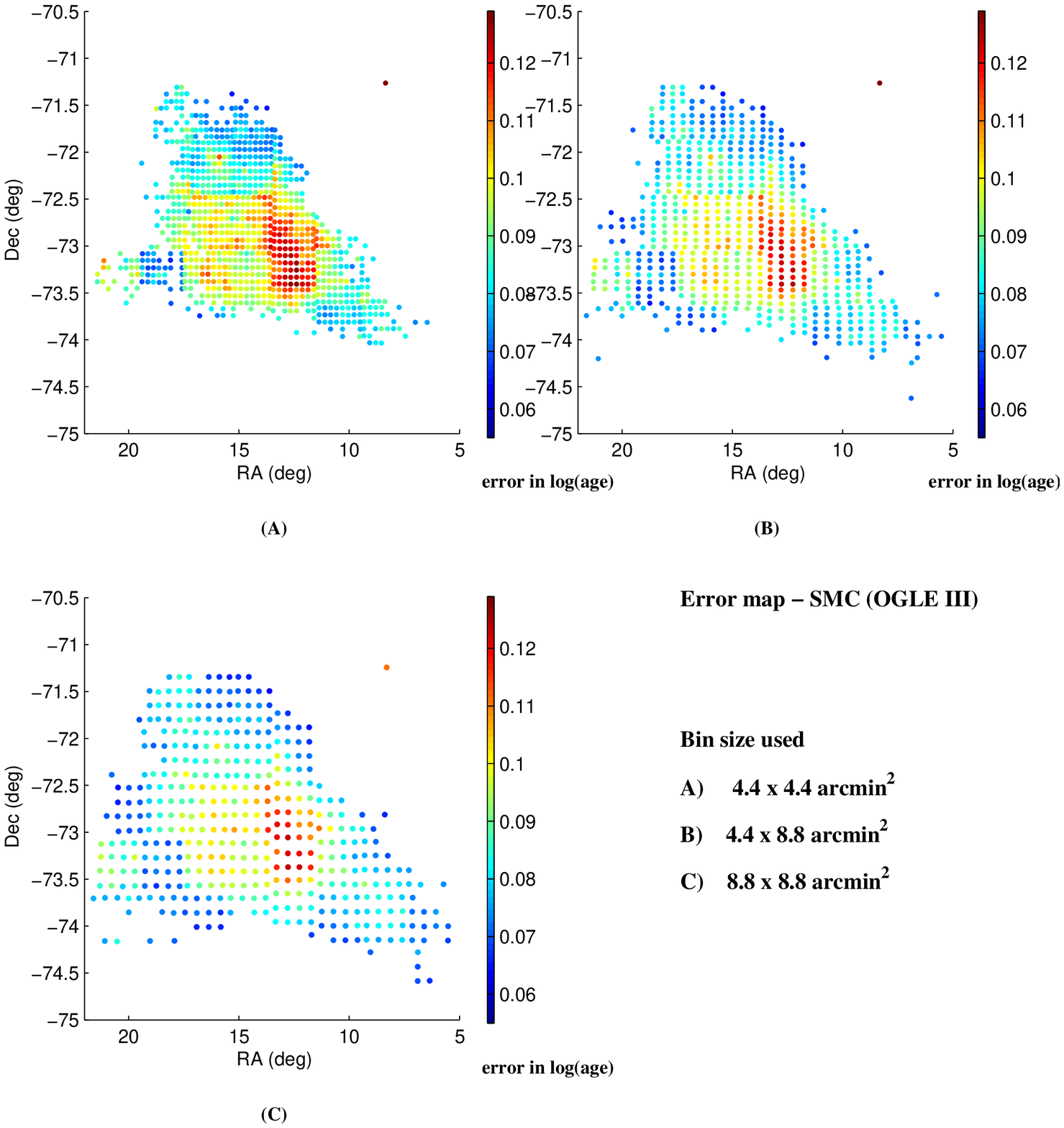}}
   \caption{Error map for the SMC using OGLE III, similar to figure 28. } 
%\\end{minipage}
    \end{figure}

\begin{acknowledgements}
We are grateful to the referee Dr. Antonio Aparicio for his valuable remarks and suggestions, which improved the presentation of the results of the paper. We gratefully acknowledge Gurtina Besla and Dr. Eva K. Grebel for their comments on the paper. We thank Dr. Katharina Glatt for providing the star cluster data and Dr. Snezana Stanimirovic for providing the HI data in the SMC. Thanks to Smitha Subramanian for interesting discussions and Dr. Stalin and Dr. Sutaria for valuable inputs.
\end{acknowledgements}


\begin{thebibliography}{}
\bibitem[Alcock et al 1999]{a99} Alcock, C., Allsman, R. A. 1999, ApJ, 117, 920
\bibitem[Bekki \& Chiba 2009]{bc86}Bekki, K., Chiba, M. 2009, PASA, 29, 48
\bibitem[Besla \& Kallivayalil 2007]{bk07}Besla, G., Kallivayalil, N. 2007, ApJ, 668, 949
\bibitem[Besla et al 2010]{b10}Besla, G., Kallivayalil, N., Hernquist, L. 2010, ApJ, 721, 97
\bibitem[Bica et al 1992]{b92}Bica, E., Claria, J.J., Dottori, H. 1992, AJ, 103, 1859
\bibitem[de Boer et al 1998]{dbr98}de Boer, K.S., Braun, J.M., Vallenary, A., \& Mebold, U. 1998, A\&A, 329, 49
\bibitem[de Vaucouleurs \& Freeman 1973]{df73}de Vaucouleurs, G., \& Freeman, K. C. 1973, Vistas Astron., 14, 163
\bibitem[Dottori \& Bica 1996]{db96}Dottori, H.; Bica, E. 1996, ApJ, 461, 742
\bibitem[Gallart et al 2008]{g08}Gallart, C. Stetson, P. B., et al 2008, AJ, 682, L89
\bibitem[Gallart et al 2009]{g09}Gallart, C., et al 2009, IAUS, 256, 281
\bibitem[Glatt et al 2010]{g10}Glatt, K., Grebel, E. K., Koch, A. 2010, A\&A, 517, 50
\bibitem[Harris \& Zaritsky 2004]{hz04}Harris, J. \& Zaritsky D. 2004, AJ, 127, 1531
\bibitem[Harris \& Zaritsky 2009]{hz09}Harris, J. \& Zaritsky D. 2009, ApJ, 138, 1243
\bibitem[Holtzman et al 1999]{h99}Holtzman, J.A., et al. 1999, AJ, 118, 2262
\bibitem[Kim et al 1998]{k98}Kim, S., Staveley-Smith, L., Dopita, M.A., Freeman, K.C., Sault, R. J., 
Kesteven, M.J., \& McConnell,D. 1998, ApJ, 503, 674
\bibitem[Kim et al 1999]{k99}Kim, S., Dopita, M.A., Staveley-Smith, L., Bessell, M. 1999, AJ, 118, 2797
\bibitem[Kim et al 2007]{k07}Kim, S., Rosolowsky, E., Lee, Y. 2007, ApJS, 171, 419
\bibitem[Lin, Jones \& Klemola 1995]{ljk95}Lin, D. N. C., Jones, B. F., \& Klemola, A. R. 1995, ApJ, 439, 652
\bibitem[Marigo et al 2008]{m08}Marigo et al. 2008, A\&A 482, 883 
\bibitem[Meaburn 1980]{m80}Meaburn, J. 1980, MNRAS, 192,365
\bibitem[Nikolaev et al 2004]{n04} Nikolaev, S., Drake, A.J., Keller, S.C., et al 2004, ApJ, 601, 260
%\bibitem[Noel et al 2007]{n07}Noel, N. E. D., \& Gallart, C. 2007, ApJ, 665, L23
\bibitem[Noel et al 2009]{n09}Noel, N. E. D. et al 2009, AJ, 705, 1260 
\bibitem[Piatek et al 2008]{p08}Piatek, S., Pryor, C., \& Olszewski, E. W. 2008, AJ, 135, 1024 
\bibitem[Pietrzynski \& Udalski 2000]{pu00}Pietrzynski, G., Udalski, A. 2000, AcA, 50, 337 
\bibitem[Saha et al 2010]{s10}Saha, A. et al 2010, AJ, 140, 1719
\bibitem[Stanimirovic et al 2004]{st04}Stanimirovic, S., Staveley­Smith, L., \& Jones, P. A. 2004, ApJ, 604, 176,186
\bibitem[Stanimirovic et al 1999]{st99}Stanimirovic, S., Staveley-Smith, L., Dickey, J. M., Sault,
R. J., \&  Snowden, S. L. 1999, MNRAS, 302, 417
\bibitem[Subramaniam 2004]{s04}Subramaniam, A. 2004, A\&A, 425, 837
%\bibitem[Subramaniam \& Subramanian 2009]{ss09apj}Subramaniam, A., \& Subramanian, S., 2009, ApJ, 703, L37
%\bibitem[Subramaniam et al 2009]{s09}Subramanian, A., \& Subramaniam, S., 2009, A\&A, 503, 
%\bibitem[Subramanian \& Subramaniam 2009]{ss09}Subramanian, S., \& Subramaniam, A., 2009, A\&A, 496, 399
\bibitem[Udalski et al.2008]{u08} Udalski, A., Soszynski, I., Szymanski, M. 2008, Acta Astron., 58, 89 (LMC OGLE3 data)
%\bibitem[van den Bergh 1999]{v99} van den Bergh, S. 1999
\bibitem[van der Marel \& Cioni 2001]{vc01}van der Marel, R.P., Cioni, M.L., 2001, AJ, 122, 1807
\bibitem[van der Marel 2001]{v01}van der Marel, R.P. 2001, AJ, 122, 1827
\bibitem[Vieira et al 2010]{vk10}Vieira, K., Girard, M., T., van Altena, F., W.,Zacharias, N., Casetti-Dinescu, I., D. 2010, AJ, 140, 1934
\bibitem[Weinberg \& Nikolaev 2001]{wn01}Weinberg, M.D., \& Nikolaev, S. 2001, ApJ, 548, 712
\bibitem[Westerlund 1997]{wes97}Westerlund, B.E., 1997, The Magellanic Clouds, Cambridge: Cambridge Univ.Press
\bibitem[Zaritsky \& Harris 2004]{zh04}Zaritsky, D., Harris, J., Thompson, I. B., et al. 2004, AJ, 123, 855 
\bibitem[Zaritsky et al 2002]{z02}Zaritsky, D., Harris, J., Thompson, I. B., Grebel, E. K., \& Massey, P. 2002, AJ, 123, 855
\bibitem[Zaritsky et al 2004]{z04}Zaritsky, D., Harris, J., Thompson, I. B., et al. 2004, AJ, 128, 1606 (LMC MCPS data)

\end{thebibliography}
\end{document}